\documentstyle[12pt]{article}
\newcommand {\cD}{{\cal D}}
\newcommand {\cE}{{\cal E}}

\newcommand {\cH}{{\cal H}}

\newcommand {\cK}{{\cal K}}
\newcommand {\cL}{{\cal L}}
\newcommand {\cM}{{\cal M}}
\newcommand {\cN}{{\cal N}}

\newcommand {\cP}{{\cal P}}
\newcommand {\cQ}{{\cal Q}}
\newcommand {\cR}{{\cal R}}

\newcommand {\cV}{{\cal V}}

\newcommand{\bA}{{\bf A}}
\newcommand{\bB}{{\bf B}}

\newcommand{\bD}{{\bf D}}

\def\a{\alpha}
\def\b{\beta}
\def\c{\chi}
\def\d{\delta}
\def\e{\epsilon}
\def\f{\phi}
\def\g{\gamma}
\def\G{\Gamma}

\def\j{\psi}

\def\l{\lambda}
\def\m{\mu}
\def\n{\nu}
\def\o{\omega}
\def\p{\pi}
\def\q{\theta}
\def\r{\rho}

\def\x{\xi}
\def\z{\zeta}
\def\D{\Delta}
\def\F{\Phi}

\def\O{\Omega}

\def\S{\Sigma}

\newcommand{\sect}[1]{\setcounter{equation}{0}\section{#1}}
\renewcommand{\theequation}{\thesection.\arabic{equation}}
\newcommand{\be}{\begin{equation}}
\newcommand{\ee}{\end{equation}}
\newcommand{\bea}{\begin{eqnarray}}
\newcommand{\eea}{\end{eqnarray}}
\newcommand{\non}{\nonumber}
\newcommand{\ba}{\begin{array}}
\newcommand{\ea}{\end{array}}

\newcommand{\ad}{{\dot{\alpha}}}
\newcommand{\bd}{{\dot{\beta}}}
\newcommand{\ve}{\varepsilon}

\newcommand{\as}{{a(2)}}
\newcommand{\bs}{{b(2)}}
\newcommand{\cDB}{{\bar\cD}}
\newcommand{\cQB}{{\bar\cQ}}
\newcommand{\cPB}{{\bar\cP}}
\newcommand{\MV}{{\vec{M}}}
\newcommand{\MBV}{{\vec{\bar M}}}
\newcommand{\mb}{{\bar \mu}}
\newcommand{\ab}{{\a\b}}
\newcommand{\abd}{{\ad\bd}}
\renewcommand{\aa}{{\a\ad}}

\newcommand{\kl}{{\a(k)\ad(l)}}

\renewcommand{\ss}{{\a(s)\ad(s)}}
\newcommand{\rr}{{\a(s-1)\ad(s-1)}}
\renewcommand{\tt}{{\a(s+1)\ad(s+1)}}

\newcommand{\pa}{\partial}
\newcommand{\na}{\nabla}
\newcommand{\da}{\dagger}
\newcommand{\hf}{\frac12}
\newcommand{\A}{{\Phi_{--}^{++}}}
\newcommand{\B}{{\bar\Phi^{--}_{++}}}

\newcommand{\os}{osp(2|4)}
\newcommand{\Os}{Osp(2|4)}
\newcommand{\osp}{osp(1|4)}
\newcommand{\Osp}{Osp(1|4)}
\newcommand{\MM}{Osp(2|4)/\cH}
\newcommand{\ra}{\rightarrow}

\def\chs{{\check \Sigma}}
\def\xb{{\bar \xi}}
\def\eb{{\bar \epsilon}}
\def\etab{{\bar \eta}}
\def\zb{{\bar \zeta}}

\def\M4{M^{7|4}}
\def\M8{M^{7|8}}

\begin{document}

\thispagestyle{empty}

\begin{flushright}
\vspace{1mm}
{March 1999}\\
\vspace{1mm}
{hep-th/9903122}\\
\end{flushright}
\vspace{1cm}

\begin{center}
{\large\bf
Explicit $N=2$ supersymmetry for higher-spin massless fields in $D=4$ AdS
	superspace.  }

\vglue 1.5 true cm

{\bf A.Yu. Segal$^{\dag}$,}

\vglue 0.3 true cm

I.E.Tamm Department of Theoretical Physics, Lebedev Physical Institute,\\
Leninsky prospect 53, 117924, Moscow, Russia

\vglue 1 true cm

{\bf A.G. Sibiryakov$^{\ddag}$,}

\vglue 0.3 true cm

Department of Theoretical Physics,\\
Tomsk State University, 634050, Tomsk, Russia
\medskip
\end{center}

\vglue 3 true cm

\begin{abstract}
\baselineskip .4 true cm
\noindent
We develop two $N=2$ superfield formulations of free equations of motion for
the joint model of all $D=4$ massless higher-superspin fields in generating
form. The explicit $Osp(2|4)$ supersymmetry is achieved without
exploiting the harmonic superspace, and with adding no auxiliary component
fields to those of $N=1$ superfields. The formulations are developed in
two different $Osp(2|4)$ homogeneous superspaces which have a
structure of a fibre bundle over the standard $D=4$ AdS superspace, with
dimensions $(7|4)$ and $(7|8)$. The $N=2$ covariant derivatives in
these spaces are expressed in terms of $N=1$ ones which gives simple
rules for component analysis.

\end{abstract}

\vfill\noindent
\dag ~on leave of absence from: Department of Physics, Tomsk State
University, Tomsk 634050, Russia;  e-mail :  {\bf segal@td.lpi.ac.ru}

\vfill\noindent
\ddag ~e-mail: {\bf sib@phys.tsu.ru}


\setcounter{section}{-1}
\section{Introduction. \label{s0} }

Not long ago, a universal formulation for the linearized dynamics of infinite
system of higher spin massless fields on anti-de Sitter space has been
suggested \cite{gks}, where the gauge-invariant action is formulated
in terms of two real and one complex {\it unconstrained} superfields on
${\tilde M}^{7|4} = \mbox{super}AdS_4 \times H_3$,
where $\mbox{super}AdS_4$ is
$N=1$ anti-de
Sitter space and $H_3$ is $\bf R^{3,1}$ one-sheeted hyperboloid. The
possibility to develop such a formulation arose due to special
properties of $N=1$ $AdS$ superfields along with the special
choice of the spin  spectrum
including every $N=1$ superspin with multiplicity one.
This spectrum structure coincides with that of
the Vasiliev's higher-spin interaction theory for $D=4$ massless higher
spin fields \cite{vas}, a significant breakthrough in
the higher spin interaction problem. It is not known up to now, whether the
equations of \cite{vas} are lagrangian or not.

The latter circumstance has urged some authors to try developing
alternative approaches. In particular, many efforts were applied to
construct different generating formulations, usually using auxiliary
operators of creation and annihilation \cite{bellon}. In this respect the
formulation of \cite{gks} is distinguished as it is based on the
commutative algebra and gives, in a sense, a theory, formulated in
extended supermanifold. This property seems to be more suitable for
describing nonlinear structures.

Another advantage of the approach undertaken in \cite{gks,ks,ksads}
is the manifest
$\Osp$ covariance. Although the equations of motion of \cite{vas} have clear
gauge structure, the question of global symmetry of higher-spin fields
has never been sufficiently studied. The hypothetical symmetry of
the actions
\cite{int} under the Fradkin-Vasiliev superalgebra of higher spins \cite{fv}
has never been observed even on the free level. Moreover, it is known
that, in the
MacDowell-Mansouri approach used in \cite{int}, even the usual global
supersymmetry transformations of supergravity are very subtle.
Contrary, in the generating superfield approach the algebra of
transformations of $N=2$ supersymmetry closes off-shell \cite{gks,gks1} and (as
we prove here) without the breaking terms proportional to gauge
transformations.
Therefore, it is natural to set the question of a manifestly
$N=2$-covariant form of GKS action.

An additional motivation to develop such
a formulation comes from the study of $N=2$ superfield
theories for Poincar{\' e} super Yang-Mills and gravity
in harmonic superspace
\cite{gikos,gios},
which present themselves a new type of supergeometries which
have no analogues in $N=1$ case.
Here it is worth noting that in harmonic superspace infinitely
many auxiliary or purely gauge component
fields were introduced to achieve an explicit
$N=2$ covariance. The model considered in the present paper also
contains infinitely many component fields, because it describes
all massless superspins. Nevertheless,
 for every superspin, the number of components
is finite. Moreover, the formulation with explicit $N=2$
supersymmetry has neither auxiliary nor purely gauge components in addition
to those of $N=1$ superfield version. Besides, to our knowledge, the $N=2$
superfield models in AdS superspace have not been considered before.
The $N=2$ AdS supergeometry differs essentially from the flat one because
$\os$ algebra does not possess the internal $su(2)$ subalgebra (rotating
supersymmetry generators) crucial for the idea
of harmonic superspace. Then, the known superfield formulations of $N=2$
supergravity contain nonminimal formulation of $N=1$ supergravity
(see also the case of superspin 3/2 in \cite{gks1}). But the description of AdS superspace
in the nonminimal formulation is unknown (see e.g. \cite{bk}). This can cause problems
in a description of $N=2$ AdS supergeometry
in the framework of the conventional approach.

On may expect that the manifest $N=2$ supersymmetry property may help
to uncover the geometry of higher-spin gauge symmetries which is
crucial for the nonlinear action construction.
That is why the superfield formulation developed in the present paper
may be taken as a starting point
to analyze the interaction problem and to construct a consistent
nonlinear action.

In this paper, we construct a manifestly $N=2$ covariant formulation for
{\it the equations of motion} of GKS theory \cite{gks}.
We construct the $N=2$ superspaces
$M^{7|4} =Osp(2|4)/\cH$
(where $\cH$ is the subgroup of $\Os$ corresponding to the
subalgebra $su(2,0|1,0) \subset \os$)
and $M^{7|8}= Osp(2|4)/SU(2) \oplus U(1)$ (differing by the number of odd
coordinates) and present two equivalent forms of GKS equations written in
terms of superfields on these superspaces.

The paper is organized as follows. In Section \ref{s1}, we describe
the $N=2$ AdS superalgebra $osp(2|4)$ and its subalgebras
$su(2,0|1,0)$
and
$su(2) \oplus u(1)$ and exhibit the decomposition of $osp(2|4)$ w.r.t. these
subalgebras. In Section \ref{s2}, we recall the GKS theory and reformulate it
by a twist which makes the $N=2$ transformations being realized by local
vector fields on $M^{7|4}$, we also observe that the $N=2$ transformation laws
may be improved by new terms, proportional to the equations of motion, what
provides the fields transformations with a more transparent form.
Remarkably, it appears that such transformations preserve just the GKS
equations of motion and not the action. But it is this form of $N=2$
transformations which is derived from the manifestly $N=2$-covariant formalism
introduced in the subsequent sections.
In Section \ref{s3}, the covariant
derivatives on homogeneous spaces ${\bar M}^{7|4} =Osp(1|4)/SU(2)$,
$M^{7|4} = Osp(2|4)/\cH$
and $M^{7|8}= Osp(2|4)/SU(2) \oplus U(1)$ are
constructed in terms of covariant derivatives on $N=1$ $AdS$ superspace
$\mbox{super}AdS_4 = Osp(1|4)/SO(3,1)$ and the so called "small vielbein" field. It
gives a possibility to study various $N=2$ superfields and equations. In Section
\ref{s4}, the special "strongly chiral" constrained superfields on $M^{7|8}$ are
studied and their component content is exhibited in terms of $N=1$
${\tilde M}^{7|4}$ superfields. In Section \ref{s5}, the GKS equations are
equivalently reformulated in terms of two strongly chiral fields on
$M^{7|8}$. This gives the anticipated manifestly $N=2$ covariant formulation for
the GKS equations of motion. In Section \ref{s6}, we show that all the stuff may
be reformulated once again in terms of superfields on $M^{7|4}$.
The obtained equations of motion describe dynamics of superfields with
indices, transforming under a local {\it super}group.
In Conclusion, we discuss the results. We supply our paper with a few
appendices carrying the information of technical character to be referred in
the main text. Appendix A contains general facts about covariant
derivatives on homogeneous spaces. The short Appendix B
introduces our supertensor notation. Appendix C describes the
$A(1|0)$ superalgebra and a brief study of its finite-dimensional
representations properties.

Let us describe our main notation.
The letters $\a ,\b , \ldots \ad ,\bd \ldots $ describe $SL(2,C)$ spinor
undotted and dotted indices. The letters $a,b,c,\ldots$ describe $su(2)$
spinor indices or $SO(3,1)$ vector ones, depending on a situation.
The letters $\x, \eta ,\ldots, \xb , \etab \ldots $ describe the indices of
two conjugated $A(1|0)$ $(1|2)-$ dimensional representations.

We consider only Lorentz tensors
symmetric in their undotted indices and separately in dotted ones.
A tensor of type $(k,l)$ with $k$ undotted and $l$ dotted indices
can be equivalently represented as
$\psi(k,l) \equiv \psi_\kl   \equiv \psi_{\a_1 \ldots
\a_k\ad_1\ldots \ad_l} = \psi_{(\a_1 \cdots \a_k)(\ad_1\cdots
\ad_l)}$.
The indices,
which are denoted by one and the same letter,
should be symmetrized separately with respect to upper and
lower indices; after the symmetrization, the maximal possible number
of the upper and lower indices denoted by the same letter
are to be contracted. In particular
$\phi_{\a(k)} \psi_{\a(l)}
\equiv \phi_{(\a_1\cdots\a_k} \psi_{\a_{k+1}\cdots\a_{k+l})}$
and $\xi^\a \phi_{\a(k)} \equiv  \xi^\b \phi_{(\b\a_1\cdots\a_{k-1})}$.
The similar notation, with a proper account of grading, takes place in the
case of supertensors, see Appendix B.  The two-dimensional antisymmetric
$\varepsilon$ tensors are defined such that $\varepsilon^{12} =
-\varepsilon_{12} = 1$.

The notation $(P|Q)$ stands for the dimension of a superspace with $P$ even and
$Q$ odd coordinates. $[A,B]= AB - (-)^{\ve (A) \ve (B)} BA$ stands for graded
commutator while $[A,B]_{-} = AB - BA$ is an ordinary commutator.

\section{$N =2$ AdS superalgebra $osp(2|4)$ and its subalgebras.\label{s1}}

In this section, we provide the information about the structure of
$osp(2|4)$ superalgebra and its decompositions with
respect to some subalgebras, this information is being used extensively
below.

The $N =2$ AdS superalgebra $osp(2|4)$ is defined via the following
graded commutation relations
\be  \label{11}
[{D ^i}_{A} , {D ^j}_{B}]  = \d^{ij} \S_{AB} - i
\ve^{ij} C_{AB} \G
\ee
\be \label{12}
[\G , {D^i}_A] = -i \ve ^{ij}  \d_{jk} {D^k}_{A}
\ee
\be \label{13}
[ \S_{AB} , {D ^i}_{C} ] = C_{CA} {D ^i}_{B} +  C_{CB} {D ^i}_{A}
\ee
\be \label{14}
[ \S_{AB} , \S_{CD} ] = C_{CA} \S_{BD} + C_{CB} \S_{AD} + C_{DA}
\S_{BC} + C_{DB}\S_{AC}
\ee
\be \label{15}
[ \S_{AB} , \Gamma ] = 0 ,
\ee
Here  $A,B, \dots = 1,2,3,4$ are $so(3,2)$ spinor representation indices;
$i,j, \dots = 1,2 $ are $so(2)$ vector ones; Kronecker delta-symbol
$\d^{ij} = \d_{ij}$ and antisymmetric tensor $\ve^{ij} = -\ve_{ij}$
are two-dimensional $SO(2)$-invariant tensors.
The matrix $C_{AB}$, $C_{AB} = - C_{BA}$, $C^2 = -1$ is $so(3,2)$ charge
conjugation matrix, it's explicit form is exhibited below
(Eq.(\ref{110'}-\ref{110ad})). The
$so(3,2)$ generators $ \S_{AB} $ are symmetric in spinor indices, $\S_{AB} =
\S_{BA}$.  The generators $ \S_{AB}$ and $\G$ are even while ${D ^i}_{A}$ are
odd.

We are interested in the decomposition of $osp(2|4)$
with respect to the maximal compact subalgebra of its even part
$su(2) \oplus so(2) \oplus so(2)$, where $su(2) \oplus so(2)$ is the
maximal compact subalgebra of $so(3,2)$ and one more $so(2)$ is
associated to the $\G$ generator.
The decomposition is performed quite simply by employing the fact
that the spinor representation of $so(3,2)$ is decomposed into the direct
sum $D(\frac{1}{2})_{+} \oplus D(\frac{1}{2})_{-}$ w.r.t. $su(2)\oplus so(2)$
subalgebra, the "$+$", "$ - $" subscripts denote the $so(2)$
weights.  Therefore, the spinor generators ${D ^i}_{A}$ are classified by
their $su(2)\oplus so(2)$ indices and by additional $so(2)$ weight associated
to the $\G$ generator:
\be \label{19}
{D ^i}_{A} = \{ D^{+}_{+a}, D^{-}_{-a},
D^{+}_{-a}, D^{-}_{+a} \} ,
\ee
where the upper index labels the
$\G$-weight:
\be \label{1-9}
{D ^{\pm}}_{A} = \frac{1}{\sqrt{2}}( {D ^1}_{A} \pm i {D ^2}_{A})
\ee
and the lower indices correspond to $su(2) \oplus so(2)$ representation,
$a,b,\dots$ are the $su(2)$ spinor indices.

According to (\ref{11}), the above classification induces the decomposition of
the even generators into subsets
\bea \label{110}
\S_{++(ab)} = [ D^{+}_{+a} , D^{-}_{+b} ]\;, \;
\S_{--(ab)} = [ D^{+}_{-a} , D^{-}_{-b} ]\\
\label{110ad} \S_{+-ab} = [ D^{+}_{+a} , D^{-}_{-b} ]
\equiv S_{(ab)} + \ve_{ab} (\G - \frac{1}{2} S)\\
\S_{-+ab} = [ D^{+}_{-a} , D^{-}_{+b} ]
\equiv S_{(ab)} + \ve_{ab} (\G + \frac{1}{2} S).
\label{1-10}
\eea
To establish the
presice form of the rest $osp(2|4)$ commutation relations in this basis we
need to know how the matrices $C_{AB}, \d^{ij}, \ve^{ij}$ look like. The
answer is
\bea \label{110'}
C_{+a\,+b} = C_{-a\,-b} =0 \\
C_{+a\,-b} = C_{-a\,+b} = - \ve_{ab} \\
\ve^{+-} =- i\; ; \;\ve^{++} = \ve^{--} =0 \\
\d^{+-} =1 \; ;\; \d^{++} = \d^{--} = 0 ,
\eea
and, therefore, the  whole system of $osp(2|4)$ commutators consists of Eqs.
(\ref{110}--\ref{1-10}) and
\be \label{13c}
[S_{a(2)},S_{b(2)} ] = 2 \ve_{ab} S_{ab} \ee
\be
\ba{cc} \label{14c}
[ S_\as, D^{\pm}_{\pm a} ] = \ve_{ab} D^{\pm}_{\pm a} &
[ S_\as, D^{\pm}_{\mp a} ] = \ve_{ab} D^{\pm}_{\mp a}
\ea \ee
\be
\ba{cc} \label{15c}
[ S, D^{\pm}_{\pm a} ] = \pm D^{\pm}_{\pm a} ; &
[ S, D^{\mp}_{\pm a} ] = \pm D^{\mp}_{\pm a}
\ea \ee
\be
\ba{cc} \label{16c}
[ \G, D^{\pm}_{\pm a} ] = \pm D^{\pm}_{\pm a} ; &
[ \G, D^{\pm}_{\mp a} ] = \pm D^{\pm}_{\mp a}
\ea \ee
\be \label{17c}
[\S_{++a(2)} , \S_{--b(2)}] = 4 \ve_{ab} (S_{ab} - \frac{1}{2} \ve_{ab} S)
\ee
\be
\ba{cc} \label{18c}
[\S_{\pm\pm a(2)} , D^+_{\mp b}] = 2 \ve_{ab} D^+_{\pm b};&
[\S_{\pm\pm a(2)} , D^-_{\mp b}] = 2 \ve_{ab} D^-_{\pm b}.
\ea \ee
The other commutators vanish. The $su(2) \oplus so(2) \oplus so(2)$
subalgebra is spanned by generators $S_{ab} \;(S_{ab} = S_{ba}), S, \G $,
where
$S_{ab}$ are $su(2)$ generators and $S, \G$ are $so(2)$ ones.

In what follows, we use the usual complex conjugation notations adopted
in supersymmetry literature (see e.g. \cite{wessbagger}):
complex conjugation does not transpose operators but
gives the additional minus sign for a product of odd operators:
$$
(D_I D_J)^\da = (-)^{IJ} D_I^\da D_J^\da.
$$
In this notation, the complex conjugation rules consistent with commutation
relations (\ref{110}-\ref{18c}) look like
\be \label{112}
 S_{a(2)}^\da = S^{a(2)} \;,\;   S^\da= -S \;,\; \G^\da =- \G,
\ee
\be
\S_{\pm\pm\as}^\da =- \S_{\mp\mp}^{a(2)}
\ee
\be  \label{14'}
(D^{\pm a}_+)^\da = - D^{\mp}_{-a},\qquad
(D^{\pm a}_-)^\da = D^{\mp}_{+a}.
\ee
This complex conjugation extracts $osp(2|4)$ superalgebra from its
complexification.

Now it is seen that
the complexification of superalgebra
$osp(2|4)$
contains two $A(1|0)$ subalgebras (see Appendix C
, Eq. (\ref{a21})) spanned by
\be \label{113}
S_{ab}\;,\; T= \G - \frac{1}{2} S \;,\; D_{+a} =D^{+}_{+a} \;,\;
D_{-a} = D^{-}_{-a}
\ee
\be  \label{15'}
S_{ab}\;,\; \tilde{T}= \G + \frac{1}{2} S \;,\; D_{+a} =D^{+}_{-a} \;,\;
D_{-a} = D^{-}_{+a} ,
\ee
which intersect over $S_{ab}$ generators.

These subalgebras are invariant w.r.t. conjugation
(\ref{112}--\ref{14'}) that extracts their
$su(2,0|1,0)$ real forms
with the even part being
isomorphic to $su(2) \oplus so(2)$.  Also, they are connected to each
other by $osp(2|4)$ automorphism $ \G' = -\G, \S'_{AB} = \S_{AB},{D
^{\pm}}'_{A} =  {D ^{\mp}}_{A}, $ therefore we can choose one of them
without an information loss.

Let's choose the subalgebra (\ref{113}) and study the decomposition of
$osp(2|4)$ adjoint representation w.r.t. it. One finds a direct sum of
graded subspaces: the first one is $su(2,0|1,0)$ itself
\be \label{114}
\cH_0 =\left
\{S_{ab}\;,\; \G - \frac{1}{2} S \;,\; D^{+}_{+a} \;,\; D^{-}_{-a}
\right\},
\ee
the second and the third ones are mutually conjugated subpaces
\be \label{115}
\ba{c}
\check\cH =\left\{ \S_{++(ab)}\;,\; D^{-}_{+a} \right\} \\ \\
\bar{\cH} =\left\{ \S_{--(ab)}\;,\; D^{+}_{-a} \right\},
\ea
\ee
eventually, the forth selfconjugated subspace is $(1|0)$ dimensional:
\be \label{116}
\tilde{\cH} = \left\{ D \equiv \frac{1}{2} (\G - S) \right\}.
\ee
In Appendix C, the finite-dimensional representations (fdrs) of
$su(2,0|1,0)$ are studied to some depth and two conjugated
$(1|2)$-dimensional "minimal" representations $\phi_{\x}$ and
$\phi_{\xb}$ are introduced (Eqs. (\ref{a22}-\ref{a24})).
Their tensor products are useful
for the description of various fdrs. In particular, the subspaces $\cH_0,
\check\cH,
\bar{\cH}, \tilde{\cH}$ transform under $su(2,0|1,0)$ according to
\be \label{117}
\ba{ccc}
\cH_0 = \cE_{\eb \z} = & \left( \ba{cc}
                                S_{ab}-\ve_{ab} T & -D_{+a}\\
                                D_{-a}         & 2T
                                \ea \right), & C^{\z \eb}  \cE_{\eb \z} =0 ,
\ea
\ee

\be \label{118}
\ba{ccc}
\check\cH =D_{\e \z} = & \left( \ba{cc}
                               \S_{--ab} & -D_{-a}^+\\
                                D_{-a}^+         & 0
                                \ea \right), &  D_{\e \z} =-(-)^{\e \z}
                                D_{\z \e}    ,
\ea
\ee

\be \label{119}
\ba{ccc}
\bar{\cH} = D_{\eb \zb} = & \left( \ba{cc}
                               \S_{++ab} & -D_{+a}^-\\
                                D_{+a}^-         & 0
                                \ea \right), &  D_{\eb \zb} =-(-)^{\eb \zb}
                                D_{\zb \eb} ,
\ea
\ee
and for $\tilde{\cH}$ one has the trivial representation.

The whole superalgebra $osp(2|4)$ may be rewritten in $su(2,0|1,0)$ covariant
form (the "$[=]$" symbol stands for the ordinary equality "$=$" with a proper
account of the sign factors in the r.h.s., see Appendix B):
\be \label{120}
[\cE_{\eb \z} , \cE_{\etab \q} ] \;\; [=] \;\;C_{\etab \z} \cE_{\eb \q} -
C_{\eb \q} \cE_{\etab \z}
\ee
\be \label{121}
[\cE_{\eb \z} , D_{\etab [2]} ] \;\; [=] \;\; 2 C_{\eb \z} D_{\etab [2]} +
2 C_{\etab \z} D_{\eb \etab}
\ee
\be \label{122}
[\cE_{\eb \z} , D_{\eta [2]} ] \;\; [=] \;\; - 2 C_{\eb \z} D_{\eta [2]} -
2 C_{\eb \eta} D_{\z \eta}
\ee
\be \label{123}
[\cE_{\eb \z} , D ] = 0
\ee
\be \label{124}
[D_{\eb [2]}, D_{\z [2]} ] \;\; [=] \;\; -4  C_{\eb \z} \cE_{\eb \z} -
8 C_{\eb \z} C_{\eb \z} D
\ee
\be \label{125}
\ba{cc}
[D, D_{\eb [2]} ] = D_{\eb [2]}, &
[D, D_{\e [2]} ] = -D_{\e [2]},
\ea
\ee
where the indices denoted by the same letters are to be
super{\it anti}symmetrized with sign factor (\ref{ab6}).
The other commutators vanish, like $[D_{\e [2]}, D_{\z [2]} ]=0 $ .


\sect{Different forms of generating theory \label{s2}}

   Let us recollect the superfield formulations of free dynamics of higher
spin fields in $N=1$ AdS superspace \cite{ksads,gks}.
We denote $4+4$ coordinates
in AdS superspace by $z$ so that $\F (z)$ is a superfield in this superspace.
All the necessary information about AdS superspace is contained in the superalgebra
of covariant derivatives $\cD_A$:
\be
\cD_A = \{\cD_\aa, \; \cD_\a, \; \cDB_\ad \equiv (\cD_\a)^*
\}\footnote{Here * means usual complex conjugation in Grassman algebra.},
\label{A-1}
\ee
exhibited by the expressions\footnote{Remember, that brackets denote
supercommutator. For example $[\cD_\a,\cDB_\ad] = \cD_\a\cDB_\ad +
\cDB_\ad\cD_\a$.}
\bea
[\cD_\a,\cDB_{\ad}]&=&-2i\cD_\aa, \qquad \;
[{\cD}_{\a\ad},{\cD}_{\b\bd}]=-2\bar\mu \mu
(\ve_{\a\b} \bar M_{\ad\bd} +
\ve_{\ad\bd} M_{\a\b}), \cr
[ \cD_\a,\cD_\b]&=&-4\bar\mu M_\ab, \qquad
[{\cD}_\a,{\cD}_{\b\bd}]=i\bar\mu
\ve_{\a\b} \cDB_{\bd}
\label{A0}
\eea
and their complex conjugated counterparts. Here $M_{\a(2)}$ and
$\bar M_{\ad(2)}$ are local Lorentz rotations,
e.g. $M_{\a(2)} \j_\b = \ve_{\b\a} \j_\a$. More
information about the geometry of AdS superspace can be found in
\cite{bk,is}. Here
we need only the fact that this supermanifold is the coset space
$\Osp/SO(3,1)$. The action of the $\osp$ algebra in AdS superspace can be
expressed in terms of covariant derivatives (see Eq.(\ref{A23}) below).

In \cite{ks,ksads}, there were found two
formulations for each superspin:  the "transversal" and the "longitudinal"
ones. We will use here only the transversal formulation for half-integer
superspin $s+1/2$ and the longitudinal one for integer superspin $s$,
denoting them as
$S^\bot_{s+1/2}$ and $S^\|_s$. It is this choice that is necessary for
constructing $N=2$ supersymmetry transformations \cite{gks1} and generating
formulation \cite{gks}.
The actions $S^\bot_{s+1/2}$ and $S^\|_s$ are expressed in
terms of the following dynamical superfields:
\bea
\cV^\bot_{s+1/2} &=& \{\, H(s,s), \G(s-1,s-1), \bar\G(s-1,s-1)\, \}
,\qquad s \geq 1 \label{A1} \\
\cV^\|_s &=& \{\, H'(s-1,s-1), G(s,s), \bar G(s,s)\, \}, \qquad
\qquad \;\;\; s\geq 1
\label{A2}
\eea
where $H$ and $H'$ are arbitrary real superfields while $\G$ and G are
constrained complex superfields, satisfying the following relations
\be
\cDB^{\ad} \Gamma_{\rr} = 0, \;\;\; \Longleftrightarrow\;\;\;
(\bar\cD^2  -2( s + 1)\mu) \G(s-1,s-1) = 0,
\label{A3}
\ee
\be
\cDB_{\ad} G_{\ss} = 0, \;\;\; \Longleftrightarrow\;\;\;
(\bar\cD^2 + 2s\mu) G(s,s) = 0
\label{A4}
\ee
and called transversal linear and longitudinal linear superfields,
respectively. These two types of superfields exhaust all possible
off-shell constrained superfields with given index structure in AdS
superspace. The gauge parameters are superfields of the same kind as the
dynamical fields: the action
$S^\bot_{s+1/2}$ is gauge invariant w.r.t. transformations with longitudinal
linear parameter $g(s,s)$ and the action $S^\|_s$ is gauge invariant w.r.t. the
ones with transversal linear parameter $\g(s-1,s-1)$.

To construct the generating formulation let us consider an $SO(3,1)$
four-vector, subjected to one-sheeted hyperboloid:
\[
q^aq_a=1,\;\;\;\;q^{\aa}q_{\aa} = -2.
\]
In \cite{gks}, it is proposed  to consider spin-tensors (\ref{A1},\ref{A2})
as the coefficients arising in expansion of analytic functions in power
series via the $q$-vector:
\be
\f(q) = \sum_{s=0}^{\infty} \f_{\ss} q^{\ss}, \qquad
q^{\ss} =\underbrace{ q^{\aa} \ldots q^{\aa}}_{s \;\; \mbox{\small times}}.
\label{A5}
\ee
These functions can be considered as the elements of commutative algebra
with the following composition law for basis elements:
\be
q_{\b(k)\bd(k)}q_{\a(k)\ad(k)} = \sum^{\min(k,l)}_{n=0} (-1)^n
\frac{c_l^n c_k^n}{c^n_{l+k-n+1}}
\underbrace{ \ve_{\b\a} \ldots \ve_{\b\a}}_{n \;\; \mbox{\small times}}
\underbrace{ \ve_{\bd\ad} \ldots \ve_{\bd\ad}}_{n \;\; \mbox{\small times}}
q_{\b(k-n)\a(l-n) \bd(k-n)\ad(l-n)}
\label{A5-5}
\ee
where $c_s^n = \frac{s!}{n!(s-n)!}$.
It successfully turned out to be \cite{gks} that the joint content of
superfield sets $\cV^\bot_{s+1/2}$ and $\cV^\|_s$ for all
values $s\ge 1$, plus two scalar superfields describing superspins $0$
and $1/2$, is represented by the following functions:
\bea
U(z,q) &=& \sum_{s=0}^{\infty} U_{\ss} q^{\ss}, \quad
U(s,s) = \G(s,s) + G(s,s);
\label{A6} \\
Y(z,q) &=& \sum_{s=0}^{\infty} Y_{\ss} q^{\ss}, \quad
Y(s,s) = H'(s,s) - H(s,s);
\label{A7} \\
X(z,q) &=& \sum_{s=0}^{\infty} X_{\ss} q^{\ss}, \quad
X(s,s) = (s+1) \{H'(s,s) + H(s,s)\};
\label{A8}
\eea
The gauge parameters are described by the function
\be
\l(z,q) = \sum_{s=0}^{\infty} \l_{\ss} q^{\ss}, \qquad
\l(s,s) = \g(s,s) - g(s,s).
\label{A9}
\ee
The functions (\ref{A6}--\ref{A9}) depend both on the vector $q$ and on
the coordinates of AdS superspace $z$, we will call them superfields.
Note that the dependence
of $U$, $Y$, $X$ and $\ve$ on their arguments is arbitrary (except for
analyticity in $q$ and reality of $Y$ and $X$) though superfields $G$,
$\G$, $g$ and $\g$ are linear (\ref{A3},\ref{A4}). This is due to the fact
that, in AdS superspace, the sum of transversal and longitudinal linear
spin-tensor superfields gives an arbitrary superfield \cite{gks,is}.

To derive an action for these superfields,  the
Lorentz-invariant trace is introduced in \cite{gks} to the algebra of
analytic functions of vector $q$:
\be
{\rm tr}\;\f = \f(0,0) \;\; \Longrightarrow \;\;
{\rm tr}\;(\f \cdot \j) = \sum_{s=0}^{\infty} \frac{(-1)^s}{s+1}\;
\f^{\ss} \j_{\ss},
\label{A10}
\ee
It is also convenient to use the following scalar operators:
\bea
\cQ = q^{\aa} \cD_{\a}\cDB_{\ad}, &&
\bar{\cQ} = - q^{\aa} \cDB_{\ad}\cD_{\a} = \cQ^*,
\label{A11}\\
\cP = \frac{\cD^2}{2\mb} = \frac 1{2\mb} \cD^\a \cD_\a, &&
\bar \cP = \frac{\cDB^2}{2\mu} = \cP^*. \label{A12}
\eea
These operators, quadratic in derivatives, satisfy the following relations
due to (\ref{A0})\footnote{In
this section, the local $so(3,1)$ rotations $M_{\a(2)}$ and $\bar M_{\a(2)}$ from
connection terms of covariant derivatives $\cD_A$ do not
act on the indices of vector $q^\aa$.}
\bea
\cQ \cPB &=& -(\cP - 1) \cQ  \cr
\cQ \cP &=& -(\cP+\cPB-2)\cQB + (\cP+1)\cQ
\label{A12-5}
\eea
It was noted in \cite{gks} that the real superfield $Y$ is
purely gauge and the gauge $Y=0$
can be imposed and substituted into the action. Then the joint action of
superfields of all superspins (including nongauge massless fields
of low spins $0$, $1/2$) takes a simple form:
\be
S = \hf {\rm tr}\displaystyle\int {\rm d}^8z\, E^{-1}  \Big\{
- \mu\mb X^2 + \hf X (\cQ U + \bar{\cQ} U^*)
- \hf U^* (\cP + \bar{\cP} - 2) U -
\hf U \bar{\cP} U - \hf U^* \cP U^*  \Big\},
\label{A13}
\ee
where ${\rm d}^8z\, E^{-1}$ is AdS superinvariant measure.
The action is invariant with respect to the following gauge
transformations with real gauge parameter $\ve(z,q)$:
\be
\d X = i (\bar{\cP} - \cP) \ve , \qquad
\d U = i  \bar{\cQ} \ve, \qquad
\d U^* = - i \cQ \ve,
\label{A14}
\ee
This gauge invariance can be checked straightforwardly with the use of
Eq.(\ref{A12-5}). It is remarkable that the corresponding calculation for
a separate superspin \cite{ksads} is much more complicated. $N=2$
supersymmetry transformations of Ref.\cite{gks1} can also be brought to
generating form.  For this purpose one should  express them via
the Killing scalar parameter
$t(z)$ \cite{gks}:
\be
t= \bar t, \qquad (\cD^2-4\mb)t=0, \qquad \cD_\a \cDB_\ad t = 0,
\label{A15}
\ee
containing one spinor and one scalar constants which are
exactly the parameters of transformations necessary to complete the
manifestly realized $N=1$ AdS superalgebra to the $N=2$ one:
\bea
\d_t Y &=& 0,\qquad
\d_t X = 2i [\bar{\cP}, t]\, \cR U + {\rm c.c.}, \cr
\d_t U &=& \frac i2 \m \bar{\m} [\cP + \bar{\cP}, t]\, \cR X
+ \frac i2 \m\bar{\m} (\bar\cP-1) [\cP + \bar{\cP}, t]\, \cR Y + \cr
&&+ i q^{\aa} (\cD_\a t) \cDB_{\ad} \cR (U + U^*). \label{A17}
\eea
Here the new operation arises which inverts the odd powers of $q_\aa$, $\cR
\Phi(q) = \Phi(-q)$.  Since the variation $\d_t Y$ vanishes, the other
variations $\d_t X$ and $\d_t U$ do not change in the gauge
$Y=0$.

     Note that the actions $S^\bot_{s+1/2}$ and $S^\|_s$ for higher spin
superfields contain the factor $(-1)^s$ to supply a positive
Hamiltonian \cite{ksads}. In generating formulation (\ref{A13}), this
sign alternation arises due to relation $q^\aa q_\aa = -2$ what
provides $(-1)^s$ in (\ref{A10}). Here we observe that there is another
possible generating formulation in which superfields (\ref{A1},\ref{A2}) emerge
in decomposition of functions on two-sheeted hyperboloid,
parametrized by a four-vector $r^\aa$, that can be chosen to be proportional
to $q$:
\be
r^\aa r_\aa = 2, \qquad r^a r_a = -1, \qquad q=ir.
\label{A18}
\ee
We will see that, in this formulation, the operation $\cR$ passes from the
global symmetry transformations (\ref{A17}) to the action and the
sign alternation still remains.
It is convenient here to define the new operation of complex conjugation
$\da$ which acts in just the same way
on the superfields not depending on $q$ and $r$
as the  operation $*$  used before,
while their actions on $q$ and $r$
are opposite:
\be
\begin{array}{cc}
q^* = q, & r^*=-r,\\
q^\da = - q, & r^\da=r.
\end{array}
\label{A19}
\ee
The fact that $*$ and $\da$ act in one and the same way on
$q$-independent superfields means that if an
action functional is real with respect to $*$ then it is real with respect
to $\da$ either.

The main goal of our reformulation $q \ra r$ is to choose
a more adequate manifold from the global symmetry point of view. We
will see that, in this new formulation, $\Os$ transformations are
realized by local operators without  $\cR$ operation. Moreover, the new
supermanifold parametrized by $(z,r)$ is the homogeneous superspace
$M^{7|4}=\Os/\cH$ where the supergroup $\cH$ is defined in the introduction
and in subsect 3.2. The $\Osp$-covariant form of
vector superfields that give the action of $\os$ superalgebra can be
constructed with the use of Killing vector:
\be
k_\aa(z) = \bar k_\aa, \qquad \cD_\a k_\aa = 0, \qquad
\cD^\a \cDB^\ad k_\aa = 0,
\label{A20}
\ee
and Killing scalar $t$ (\ref{A15}). It is convenient to introduce the
following notation for all linearly independent derivatives of these
Killing superfields:
\bea
k_\a = \frac i8 \cDB^\ad k_\aa, && \bar k_\ad = (k_\a)^\da, \cr
k_{\a(2)} = \cD_\a k_\a, && \bar k_{\ad(2)} = (k_{\a(2)})^\da, \label{A2-1}\\
t_\a = \hf \cD_\a t, && \bar t_\ad = (t_\a)^\da.
\label{A22}
\eea
Then the vector superfield giving the action of  $\osp$ superalgebra
can be written as usual (see e. g. \cite{bk}),
\be
\cK = -\hf k^\aa \cD_\aa + k^\a \cD_\a + \bar k_\ad \cDB^\ad
+ k^{\a(2)} M_{\a(2)} + \bar k^{\ad(2)} \bar M_{\ad(2)} =
\cK^\da = \cK^*.
\label{A23}
\ee
It appears that the rest vectors of $\os$ are parametrized
by Killing scalar~$t$,
\be
\cM = 2i r^\aa (t_\a \cDB_\ad + \bar t_\ad \cD_\a) + 2it \na,
\qquad \cM^\da = \cM = -\cM^*,
\label{A24}
\ee
where $\na$ is an imaginary vector field
\be
\na = - \frac i4 (\cQ - \cQB) = - \frac i2 r^\aa \cD_\aa
\label{A2-5}
\ee
It can be straightforwardly checked that a commutator of two operators $\cM$
closes on $\cK$,
\be
[ \cM (t), \cM (t') ] = - \cK (k), \qquad
k_\aa = 16i (t_\a \bar{t'}_\ad + \bar t_\ad t'_\a).
\label{A2-6}
\ee
The conclusion that the coordinates $(z,r)$ parametrize superspace
$\MM$ will be proved by the construction of $\Os$-covariant
derivatives with the proper algebra (see Sec. \ref{s3}). Thus, vector
superfields $\cM$ and $\cK$ give the action of $\os$ superalgebra.

Depending on $r$ are new superfields $Z = Z^\da$ and $V$ whose contents
coincide with the contents of $X$ and $U$, respectively.
\bea
Z(z,r) &=& \sum_{s=0}^\infty r^\ss Z_\ss, \qquad
           Z(s,s) = (-1)^{[\frac s2]} X(s,s), \label{A2-7}\\
V(z,r) &=& \sum_{s=0}^\infty r^\ss V_\ss, \qquad
           V(s,s) = (-1)^{[\frac{s-1}2]} U(s,s),\label{A2-8}
\eea
where $[\l]$ denotes the integral part of $\l$. Note that, even though $q$ is
connected to $r$ by (\ref{A18}), and superfields $Z$, $V$ and $X$, $U$
have the identical content, they are not proportional. Therefore, the new
formulation cannot be obtained by a trivial change like $X \ra \a Z$, ${U
\ra \b V}$, $q \ra ir$ in the action (\ref{A13}). Nevertheless,
it is not difficult to check that every term in the action (\ref{A13}) can be
equivalently rewritten in new formulation with the use of operation $\cR$:
\bea
S &=& \hf {\rm tr}\displaystyle\int {\rm d}^8z\, E^{-1}  \Big\{
- \mu\mb Z \cR Z + \frac i2 Z \cR (\cQ V + \bar{\cQ} V^\da) - \cr
&&{}- \hf V^\da (\cP + \bar{\cP} - 2) \cR V -
\hf V \bar{\cP} \cR V - \hf V^\da \cP \cR V^\da  \Big\},
\label{A2-9}
\eea
where '${\rm tr}$' is the same
operation as in (\ref{A10}): ${\rm tr}
\Phi(r) = {\rm tr} \Phi(-iq) \equiv \Phi(0,0)$,
$\cQ$ and $\bar \cQ = - \cQ^\da$ are the same operators (\ref{A11}) with $ir$
being substituted for $q$.

Let us turn to $N=2$ supersymmetry transformations. It is shown
in Ref.\cite{gks1}
that their algebra closes off-shell. Nevertheless, this algebra
is broken by the terms proportional to gauge transformations (see \cite{gks}
for the  generating form of this algebra). Here we observe that
it can be closed eventually by adding certain gauge transformation to
(\ref{A17}). The structure of the global symmetry becomes more clear after
splitting $V$ into its real and imaginary parts (with respect to $\da$)
$V= \g + i\r$.  Then the action (\ref{A2-9}) reads
\bea
S &=& \hf {\rm tr}\displaystyle\int {\rm d}^8z\, E^{-1}  \Big\{
-\mu\mb Z \cR Z - 2i Z \cR \na \r + \r \cR \r + \cr
&& + \frac i2 Z \cR (\cQ+\bar\cQ) \g + i \g \cR (\cP-\bar\cP) \r -
\g (\cP + \bar\cP -1) \cR \g \Big\}.
\label{A30}
\eea
Gauge transformations (\ref{A14}) convert into
\be
\d Z = -i (\cP-\cPB) \ve, \qquad
\d \r = - \frac i2 (\cQ+\cQB) \ve, \label{A31}\qquad
\d \g = - 2i \na \ve,
\ee
From (\ref{A30},\ref{A31}) it is seen that $Z$ and $\r$ are similar fields
because they are both auxiliary (either $Z$ or $\r$, not both, can be
excluded by the equations of motion), and enter the action in an analogous
way. One can show that the superfield $\g$ is purely gauge. Really,
using (\ref{A18})
and (\ref{A5-5})
one can derive gauge transformations for the components $\g(s,s)$ of
the superfield $\g$:
$$
\d \g_\ss = a_s \cD_\aa \ve_\rr + b_s \cD^\aa \ve_\tt
$$
where $b_s \ne 0$ and $a_s$ are numerical factors. Using the freedom in the
choice of $\ve_\tt$ we can put $\g_\ss = 0$. This gauge imposes the
differential restriction $\na \ve = 0$ on the parameter and therefore cannot
be inserted in action.

    To obtain the $N=2$ supersymmetry transformations with completely
closed algebra one should add to transformations (\ref{A17}) the gauge
variation (\ref{A31}) with the parameter $\ve= -2t \g$, rewritten in
terms of superfields $Z$, $\r$ and $\g$.  The result is
\bea
\tilde \d_t \g &=& \cM \g,   \cr
\tilde \d_t Z &=& -n\r + it(\cP-\cPB)\g, \label{A32}\\
\tilde \d_t \r &=& \mu\mb n Z + \frac i2 t(\cQ+\cQB)\g, \cr
n &=& [\cP+\cPB,t] = \cN + 4t, \qquad
\mu\mb \cN = 2 (\mu t^\a \cD_\a + \mb \bar t_\ad \cDB^\ad) \non
\eea
Here $\cM$ is a vector superfield of  $\os$ superalgebra (\ref{A24}) and
$\cN$ is another vector superfield whose commutator has the opposite,
as compared to (\ref{A2-6}), sign.
\be
[\cN(t),\cN(t')] = \frac 1{\mu\mb} \cK = [n(t),n(t')].
\label{A33}
\ee
Note that $n$-dependent terms in (\ref{A32})
may be represented with the use of matrix $J$:
\be
\left( \ba{c}
       \tilde\d Z\\
       \tilde\d \r
       \ea \right) = J n
\left( \ba{c}
       Z\\
       \r
       \ea \right) + \ldots, \qquad
J=\left(\ba{cc}
0      & -1 \\
\mu\mb &  0
\ea\right), \qquad J^2 = -\mu\mb,
\label{A34}
\ee
and the dots denote $\g$-dependent terms. This provides the correct sign
in the commutator:
\be
[\tilde\d_t, \tilde\d_t'] = -\cK,
\label{A35}
\ee
where $\cK$ is the Killing operator (\ref{A23}) with vector parameter $k_\aa$
defined in (\ref{A2-6}). The difference in sign with \cite{gks} is
conditioned by a different definition of commutator of
variations $[\tilde\d_t, \tilde\d_t']$.

Let us emphasize that the superfield $\g$ transforms via itself
by the supervector operator $\cM$. The whole representation of the
superalgebra $\os$ on superfields $\g$, $Z$ and $\r$ has a semidirect sum
structure. Surprisingly, it is possible to split it completely
by adding variations proportional to the equations of motion. Let us
write them down
\bea
{\d S}/{\d Z} &\equiv& - 2\mu\mb \cR Z - 2i\cR\na\r
+ \frac i2 \cR (\cQ+\cQB)\g = 0, \label{A361}\\
{\d S}/{\d \r} &\equiv&  2 \cR \r - 2i\cR\na Z
+ i \cR (\cP-\cPB)\g = 0, \label{A362}\\
{\d S}/{\d \g} &\equiv& \frac i2 \cR(\cQ+\cQB)Z + i\cR(\cP-\cPB)\r
-2(\cP+\cPB-1)\cR\g = 0. \label{A363}
\eea
Then we can define new variations $\bar\d_t$ with the same
algebra:
\bea
\bar\d_t \g &=& \tilde\d_t \g = \cM \g, \cr
\bar\d_t Z &=& \tilde\d_t Z - t\cR {\d S}/{\d \r}
= -(\cN + 6t)\r + 2it\na Z, \label{A37}\\
\bar\d_t \r &=& \tilde\d_t \r - t\cR {\d S}/{\d Z}
= \mu\mb(\cN + 6t)Z + 2it\na \r. \non
\eea
Unfortunately, the new transformations leave invariant only the equations of
motion (\ref{A361}--\ref{A363}) and not the action (\ref{A2-9}). The reason is
that to maintain the invariance of the action, the expression
$\bar\d - \tilde\d$ to transformations should be a composition of
an antisymmetric matrix with the action variations. However, in
(\ref{A37}) a symmetric matrix is used.

\sect{Structure of supercovariant derivatives.\label{s3}}

Now we begin the description of the formalism with explicit $\Os$
symmetry. To develop it we need the $\Os$-covariant derivatives and their
expressions in terms of $\Osp$-covariant operators used previously.

In this section, the following sequence of derivatives (which starts from
ordinary $\Osp/SO(3,1)$ derivatives (\ref{A-1}) defined in $N=1$ AdS
superspace) is constructed: $\Osp/SU(2)$, $\MM$ and $\Os/(SU(2)\oplus
U(1)) \equiv \Os/U(2)$. Two latter sets of $\Os$-covariant derivatives
will be used in the explicitly $N=2$ invariant formulation of higher-spin
equations.  We will construct every set in this sequence in terms of the
preceding one. In this way, all the expressions for covariant derivatives
may be expressed in terms of $\Osp/SO(3,1)$ ones, that provides us with
the possibility to make a bridge between various $N=2$ covariant equations
and those of $N=1$ GKS theory.

Let us sketch our technology.  Recall that, according to general facts
collected in App.A, the derivatives covariant with respect to a
(super)group $G$ are standardly determined for every coset space $G/H$
where $H$ is the local subgroup. In particular, given a supergroup $G$ of
dimension $(P|Q)$ and a supergroup H of dimension $(p|q)$, there exist
$(P-p|Q-q)$ covariant derivatives for $G/H$ coset space, and the rest
$(p|q)$ generators span a local $H$ supergroup.

We will start with $(4|4)$ $\Osp/SO(3,1)$ covariant derivatives in $N=1$
AdS space $M^{4|4}$.  Then we will observe that the extension of $M^{4|4}$
by a constrained Lorentz four vector $r^\aa$ (being the essence of GKS
formulation) can be considered as the $(7|4)$-dimensional homogeneous
space $\Osp/SU(2)$, and build corresponding $(7|4)$ covariant derivatives.
To this aim, we introduce a "small vielbein field" (\ref{d5}, \ref{d6})
and construct the $\bar M^{7|4}$ derivatives via $M^{4|4}$ ones and the
small vielbein (subsection 3.1).  Further, the $\Os/\cH$ superspace has
the same dimension $(7|4)$ as the superspace $\Osp/SU(2)$, and the numbers
of covariant derivatives in these spaces are equal.  In fact, these new
$N=2$ covariant derivatives in a certain basis are given by $N=1$ ones,
this is shown in subsection 3.2.  It will remain to build the $M^{7|8}$
superspace (subsection 3.3), which differs from two previous ones by the
number of odd coordinates.  We will not inrtroduce these new odd
coordinates $\theta$ explicitly, instead, we describe an $M^{7|8}$
superfield by its $M^{7|4}$ proections, associated with $\theta=0$
components of its covariant derivatives.

Before starting the superspace building process let us make a remark.
Note that the supergroup $\cH$ serves as a local supegroup  for
$\MM$-covariant derivatives. The corresponding superfields in the coset
space carry the indices which transform under the supergroup (we call
superindices the indices of such a kind).  Therefore these indices are
graded, they have even and odd parts.  For example, one has the superindex
$\xi$ (\ref{a22}):  $\f_\xi = (\f_{+a},\f_{++})$. The correspondinng
superfield has the odd part $\f_{+a}$ and the even part $\f_{++}$ (and
each of them has even and odd components after an expansion in power
series by odd coordinates).  As far as we know superfields with
superindices have not yet been used in physically interesting models. In
Sect.\ref{s6}, we present the models with these superfields.

{\bf 3.1 $\Osp/SU(2)$-covariant derivatives.}

\noindent Consider the supermanifold $\bar M^{7|4}$ parametrized by coordinates
$(z,r)$ defined in Sect.2.  The dynamical superfields of the theory under
consideration are $r$-analytic functions in this supermanifold, (\ref{A2-7},\ref{A2-8}).
The supergroup $\Osp$ acts on $\bar M^{7|4}$
transitively. Indeed, it acts transitively on the
supermanifold $M^{4|4}$ parametrized by $z$, then, the stability subgroup for
$z$ -- the Lorentz group $SL(2,C)$ -- acts transitively on a two-sheeted
hyperboloid parametrized by $r$. The stability subgroup of the point
$(z,r) \in \bar M^{7|4}$ is $SU(2)$. According to
App. A, there exist $\Osp/SU(2)$ covariant derivatives in this
superspace $\bar M^{7|4}$. In this subsection, we find their expressions in
terms of covariant derivatives (\ref{A0}) of superspace $M^{4|4}$.
Our guide is the requirement for $\Osp/SU(2)$ covariant derivatives, together with
local generators of $su(2)$, to form the superalgebra $\osp$. It is
convenient to introduce $su(2)$-covariant basis in this superalgebra
which has the following structure
\be
S_{a(2)},\qquad \S_I = \{ \S_{\pm\pm\as}, \S_{\pm a}, \S \}.
\label{d1}
\ee
The superalgebra $\osp$ is defined by the following commutation relations
equivalent to (\ref{A0}):
\bea
[S_\as, S_\bs] &=& 2 \ve_{ab} S_{ab},   \label{d2-1}\\
{}[S_\as, \S_{\pm\pm b(2)}] &=& 2\ve_{ab} \S_{\pm\pm ab}, \qquad
     [S_\as, \S_{\pm b}] = \ve_{ab} \S_{\pm a},  \label{d2-2}\\
{}[\S, \S_{\pm\pm\as}] &=& \pm \S_{\pm\pm\as}, \qquad
     [\S, \S_{\pm a}] = \pm \hf \S_{\pm a}, \label{d2-3}\\
{}[\S_{++\as}, \S_{--b(2)}] &=& 4\ve_{ab}
     (S_{ab} - \ve_{ab} \S), \label{d2-4}\\
{}[\S_{\pm\pm\as}, \S_{\mp b}] &=& 2\ve_{ab} \S_{\pm a}, \label{d2-5}\\
{}\{ \S_{\pm a}, \S_{\pm a} \} &=& - \S_{\pm\pm\as}, \qquad
     \{ \S_{\pm a}, \S_{\mp b} \} = - S_{ab} \pm \ve_{ab} \S. \label{d2-6}
\eea
with reality conditions
\bea
&(S_\as)^\da = S^\as&
\label{d3}     \\
&(\S_{\pm\pm\as})^\da = -\S_{\pm\pm}^\as, \qquad
(\S_{\pm a})^\da = \mp \S_{\mp}^a, \qquad \S^\da = -\S&
\label{d4}\eea
Commutation relations (\ref{d2-1}), together with reality conditions
(\ref{d3}), define the $su(2)$ algebra.
The rules (\ref{d3},\ref{d4}) reflect the fact that the complex
conjugate of the fundamental representation of $su(2)$ is
equivalent to its contragredient representation.

Let us now discuss the
connection between the considered basis $\S_I$, $S_\as$
and the basis $\cD_A$, $M_{\a(2)}$, ${\bar M}_{\ad(2)}$
 (\ref{A0}) used in the previous section.
To express the former via the latter we need to pass
to $SU(2)$-covariant indices. Note that conjugation rules for generators
$\cD_A$, $M_{\a(2)}$, $\bar M_{\ad(2)}$ are as follows:
\be
(M_{\a(2)})^\da = \bar M_{\ad(2)}, \qquad (\cD_{\a\bd})^\da = \cD_{\b\ad},
\qquad (\cD_\a)^\da = \cDB_\ad
\label{d4-5}
\ee
That is why if $\cD_\a$ is transformed under the fundamental
representation of $SU(2)$, then $\cDB_\ad$ is transformed under the
contragredient one. Therefore we should define
\be
M_{(ab)} = M_{(\a\b)}, \quad \bar M^{(ab)} = \bar M_{(\ad\bd)}, \quad
\cD_{a,b} = {\cD_\a}^\bd, \quad \cD_a = \cD_\a, \quad \cDB^a = - \cDB_\ad,
\label{d4-6}
\ee
here, in every equality, $a$ and $b$ take the same values as $\a$ and $\b$
(or $\ad$ and $\bd$), respectively. Now, the  commutation relations
(\ref{d2-1}--\ref{d2-6}) are connected with (\ref{A0}) by the expressions
\bea
\S_{\pm\pm\as} &=& \cD_{a,a} \pm i(\bar M_\as - M_\as), \cr
S_\as &=& -(M_\as + \bar M_\as), \qquad \
\S = - \frac i2 {\cD^a}_{,a} \label{d4-7}\\
\S_{+a} &=& \frac {\bar j}2 (\cD_a + \cDB_a), \qquad
\S_{-a} = - \frac j2 (\cD_a - \cDB_a),\non
\eea
where $j=e^{i\pi/2}$.

The covariant derivatives (we  denote them $\check\S_I$) possessing
the required algebra and reality conditions should depend both on $z$ and $r$.
To find their expressions,
first of all we introduce a 'small vielbein' ${e_a}^\a(z,r)$ and its
conjugated $({e^a}_\a)^\da = e_{a\ad}$\footnote{As usual, we raise and
lower indices $\a$, $\ad$ and $a$ of the small vielbein with help of
antisymmetric symbols $\ve_\ab$, $\ve_\abd$, $\ve_{ab}$ and so on.  For
example, ${e^a}_\a = \ve_\ab \ve^{ab} {e_b}^\b$.} which convert Lorentz
two-component spinor indices $\a$ and $\ad$ into two-component spinor
$SU(2)$-index $a$.  We subject this vielbein to the following conditions
invariant both with respect to $SL(2,C)$ and $SU(2)$. First, the
$2\times 2$-matrix $i{e_a}^\a$ is an element of the $SL(2,C)$ group:
\be
\ve_{ab} = - {e_a}^\a {e_b}^\b \ve_\ab.
\label{d5}\ee
Second, it is convenient to restrict partially the remaining freedom by
\be
e^{a\a} {e_a}^\ad = r^\aa,
\label{d6}\ee
(remember that the vielbein $e$ is a function of $z$ and $r$). One can see
that the latter condition (\ref{d6}) is consistent with $r^\aa r_\aa = 2$
and $(r_\aa)^\da = r_\aa$. It implies
$$
{e_a}^\a {r_\a}^\ad = {e_a}^\ad, \qquad {e_a}^\ad {r_\ad}^\a = - {e_a}^\a.
$$

    Let us remind that the Lorentz generators $M_{\a(2)}$, $\bar
M_{\ad(2)}$ enter the connection parts of covariant derivatives $\cD_A$
in all formulas of Sect. \ref{s2}. These generators do
not act on the indices of vector $r^\aa$. Their action on a scalar superfield
$\F(r)$ is equivalent to the action of the following operators
\be
\MV_{\a(2)} = \hf {r_\a}^\ad \pa_\aa, \qquad
\MBV_{\ad(2)} = \hf {r_\ad}^\a \pa_\aa, \qquad
\pa_\aa \equiv \frac \pa{\pa r^\aa}
\label{d6-6}
\ee
that preserve the constraint $r^2 = - 1$ and therefore are (complex)
vector fields on the hyperboloid. Let us introduce operators $\hat
M_{\a(2)}$, $\hat{\bar M}_{\ad(2)}$ that act on all indices $\a$, $\ad$
and include the differential operator:
\bea
\hat M_{\a(2)} {e^a}_\b &\equiv& \MV_{\a(2)} {e^a}_\b +
\ve_{\b\a} {e^a}_\a, \cr
\hat M_{\a(2)} r_{\b\bd} &\equiv& \MV_{\a(2)} r_{\b\bd} +
\ve_{\b\a} r_{\a\bd} = 0, \label{d6-7}\\
\hat M_{\a(2)} \F(r) &\equiv& \MV_{\a(2)} \F(r) \non
\eea
and conjugated expressions for $\hat{\bar M}_{\ad(2)}$. Then define two sets of
modified covariant derivatives $\cD_A$, substituting either $\MV_{\a(2)}$ or
$\hat M_{\a(2)}$ instead of $M_{\a(2)}$ in the connection terms:
$$
\vec \cD_A \equiv \cD_A|_{M\ra\MV}, \qquad
\hat \cD_A \equiv \cD_A|_{M\ra\hat M}.
$$
Then we have the following equations for an arbitrary $SU(2)$ spinor superfield
$\F_a(z,r)$:
\bea
\hat \cD_A \F_a(z,r) &=& \vec \cD_A \F_a(z,r) = \cD_A \F_a(z,r),\cr
\hat \cD_A \hat \cD_B \F_a(z,r) &=& \cD_A \cD_B \F_a(z,r)
\label{d6-8}
\eea
Now we choose $(7|4)$ independent vector fields:
\bea
\na_\as &\equiv& {e_a}^\a {e_a}^\ad \vec\cD_\aa = - (\na^\as)^\da, \qquad
    \na \equiv - \frac i2 r^\aa \vec\cD_\aa = -\na^\da,    \cr
\pa_\as &\equiv& -i {e_a}^\a {e_a}^\ad \pa_\aa = -2i {e_a}^\a {e_a}^\a
\MV_{\a(2)} = (\pa^\as)^\da, \label{d7}\\
\na_a &\equiv& {e_a}^\a \vec\cD_\a, \qquad
    \bar\na_a \equiv {e_a}^\ad \vec\cDB_\ad = (\na^a)^\da  \non
\eea
It is also convenient to define the modified operators $\hat\na$:
\be
\hat\na_A \equiv \na_A|_{\MV\ra\hat M}, \quad
\hat\na_A \F_a(z,r) = \na_A \F_a(z,r),\quad
\hat\na_A \hat\na_B \F_a(z,r) = \na_A \na_B \F_a(z,r)
\label{d7-5}
\ee
Note their useful properties:
$$
\pa_\as r^\aa = 2i {e_a}^\a {e_a}^\ad,\qquad
\hat\na_A r^\aa = 0
$$
In the analogous equation for $\na_A$, the connection parts of covariant
derivatives act on $r^\aa$ and do not give zero.
Making use of these equations, all independent
first order derivatives of the small vielbein prove to be described
by the expressions
\bea
v_{\as\bs} &\equiv& \hf \{ (\pa_\as {e_b}^\a) e_{b\a} +
    (\pa_\as {e_b}^\ad) e_{b\ad} \},    \label{d8-1}\\
&& \pa_\as {e_b}^\a = - v_{\as\bs} e^{b\a} - i\ve_{ab} {e_a}^\a, \cr
u_{\as\bs} &\equiv&  (\hat\na_\as {e_b}^\a) e_{b\a} =
    (\hat\na_\as {e_b}^\ad) e_{b\ad},    \label{d8-3}\\
w_{\as} &\equiv&  (\hat\na {e_a}^\a) e_{a\a} =
    (\hat\na {e_a}^\ad) e_{a\ad},        \label{d8-4}\\
y_{a\bs} &\equiv&  (\hat\na_a {e_b}^\a) e_{b\a} =
    (\hat\na_a {e_b}^\ad) e_{b\ad}, \qquad
    \mbox{and complex conjugated.}\label{d8-5}
\eea
To derive the algebra of the vector fields $\pa_\as$, $\na_A$ it is
sufficient to calculate the algebra of the modified operators $\pa_\as$,
$\hat\na_A$ since, due to (\ref{d7-5}), these algebras coincide modulo curvature
terms with local rotations $M_{\a(2)}$. The latter algebra can be derived
straightforwardly using the commutation relations for $\hat\cD_A$
(given by (\ref{A0})) along with the substitution
$\cD_A \ra \hat\cD_A$, $M \ra \hat M$. In so doing one can deduce that the supercommutators of vector fields
$\pa_\as$, $\na_A$ close on themselves with coefficients of
non-holonomicity being
 proportional to tensor superfields $v$, $u$, $w$, and $y$.

Now we are in a position to construct the operators
\bea
\cL_\as &=& -\pa_\as - {v_\as}^\bs S_\bs, \label{d9}\\
\cD_\as &=& \na_\as + {u_\as}^\bs S_\bs,  \label{d10}\\
\check\S      &=& \na + w^\bs S_\bs,  \label{d11}\\
\cD_a   &=& i\na_a + i{y_a}^\bs S_\bs,
\label{d12}\eea
acting on sections of $su(2)$ fiber bundle over $\bar M^{7|4}$, with
$S_{a(2)}$ representing local $su(2)$ generators.
In fact, these expressions
already define $\Osp/SU(2)$-covariant derivatives, but for our purpose
it is more convenient to make a linear substitution
\bea
&\check\S_{\pm\pm\as} = \cD_\as \pm \cL_\as,& \label{d13}\\
&\check\S_{+a} = \frac{\bar j}2 (\cD_a + \cDB_a), \qquad
    \check\S_{-a} = -\frac j2 (\cD_a - \cDB_a).& \label{d14}
\eea
Using the algebra of vector fields (\ref{d7})
one can show
that the introduced operators $\check\S_I$, along with
$S_{ab}$ local rotations, obey the superalgebra
$osp(1|4)$ (\ref{d2-3}--\ref{d2-6}). They
also satisfy the reality conditions (\ref{d4}).
The obtained results exhibit a general situation, briefly discussed in App. A.
Suppose that we have two
homogeneous manifolds $G/L$ and $G/H$ for the same group $G$, and $H$ is a
subgroup of $L$,  $H \subset L \subset G$. Then there exists a general
expression for $G/H$-derivatives via $G/L$-ones.
The small vielbein field can be viewed as an object which parametrizes
different embeddings $L/H \subset L$.


{\bf 3.2 $\Os$-covariant derivatives in $M^{7|4}$.}

Here we construct other covariant derivatives in the same supermanifold
$\bar M^{7|4}$, where the derivatives $\check\S_I$ were defined
previously. The point is that, according to Sect.\ref{s1}, this
supermanifold can be considered as the homogeneous manifold $\MM$ for
$N=2$ AdS supergroup $\Os$, the corresponding vector fields spanning
$osp(2|4)$ superalgebra are given by Eqs.(\ref{A23}, \ref{A24}).  Here, as
it is mentioned in the Introduction, $\cH$ denotes the (super)subgroup of
the supergroup $\Os$, associated with superalgebra $su(2,0|1,0) \subset
\os$. Note that the bosonic part of $su(2,0|1,0)$ coincides with
$u(2)=so(3)\oplus so(2)$.  The algebra of the corresponding
$\Os$-covariant derivatives and local rotations is $\os$ superalgebra.  We
denote these $\MM$ derivatives by the same letter as the corresponding
generators in (\ref{110}--\ref{1-10},\ref{13c}--\ref{18c}) with the
additional emphasis '$\check{~}$'.  The local rotations are denoted
identically with the generators of $A(1|0)$:
\bea
\mbox{local rotations} && S_\as,\; T=\G-S/2,\; D^\pm_{\pm a} \cr
\mbox{covariant derivatives} && \check\S_{\pm\pm\as},\; \check D^\pm_{\mp a},\;
\check D=\G/2-S/2
\label{d14-5}
\eea
The commutation relations in this basis can be read off
(\ref{110}--\ref{1-10},\ref{13c}--\ref{18c}) after switching from
$\G$, $S$ to $T$, $\check D$. To derive the
$\Os$-covariant derivatives (\ref{d14-5})
in terms of derivatives $\check\S_I$ constructed
in the previous subsection, it is convenient to pass to a new
basis. Let us recall the odd generators $D^1_{\pm a}$ and
$D^2_{\pm a}$ related to $D^\pm_{\pm a}$ and $D^\pm_{\mp a}$ by the rules
(\ref{1-9}). Introduce the notation
\be
\check\S_{\mp a} = \check D^2_{\mp a} = \pm \frac 1{\sqrt{2}}
(\check D^\pm_{\mp a} - D^\mp_{\mp a}), \qquad  \check\S = T - 2\check D.
\label{d15}\ee
Strictly speaking, $\check\S_{\pm a}$ and $\check\S$ are no longer
pure covariant
derivatives for $\MM$ but the combinations of covariant derivatives and
local $\cH$ rotations. Therefore,
the supercommutator of $\check\S_{\pm a}$ with local
rotations $D^\pm_{\pm a}$ is not a linear combination of bosonic covariant
derivatives and contains explicitly local rotations $S_\as$. Nevertheless,
it is possible to consider the following basis of operators
(the set of generators $D^\pm_{\pm a}$, $D^2_{\pm a}$
forms the basis in the odd subspace of superalgebra $\os$):
\bea
\mbox{$\cH$ - local rotations} && S_\as,\; T,\; D^\pm_{\pm a} \label{d15-3}\\
\mbox{combinations} && \check\S_{\pm\pm\as},\; \check\S_{\pm a},\; \check\S
\label{d15-5}
\eea
Now our task reduces to finding the operators $\check\S_I$.
The basis (\ref{d15-3},\ref{d15-5}) for $\os$ was chosen so that
generators $\check\S_I$, $S_\as$ span $\osp$ subalgebra, which
can be derived from Eqs.(\ref{110}--\ref{1-10},\ref{13c}--\ref{18c}).
This means that the commutators of operators $\check\S_I$
close on $\check\S_I$ and $S_\as$ and coincide with those of
derivatives $\check\S_I$ explicitly constructed in the previous subsection.
That is why we can identify the desired combinations (\ref{d15-5})
with the available operators
(\ref{d13}--\ref{d14}). Thus we do not need to 'construct' any new objects
to built $\MM$ covariant derivatives.
In this reasoning, we have used the fact that the whole superalgebra $\os$
can be spanned by two intersecting subalgebras:  $\osp$ with basis $\S_I$,
$S_\as$ and subalgebra of local rotations $S_\as$, $D^\pm_{\pm a}$, $T$.
The action of $\osp$ was defined in the subsection 3.1 while the local
rotations act by definition on the indices of covariant derivatives.
Of course, we
can go back to the basis (\ref{d14-5}) that
appears to be more convenient in what
follows:
\bea
\check D^\pm_{\mp a} &=& \pm \sqrt{2} \check\S_{\mp a} + D^\mp_{\mp a},  \label{d16}\\
\check D &=& -  \check\S/2 + T/2.
\label{d17}\eea
Let us overview the subsection results. Even covariant derivatives
$\check\S_{\pm\pm\as}$ are given by (\ref{d13}).  Odd derivatives $\check
D^\pm_{\mp a}$ and scalar derivative $\check D$ are given by (\ref{d16}),
where $\check\S_{\pm a}$, $\check\S$ are $\Osp$-covariant derivatives
(\ref{d11},\ref{d14}) and $D^\pm_{\pm a}$, $T$ are local rotations. Let us
stress that here only a particular solution is found for
$\MM$-derivatives.  That is why it is not $\cH$-invariant and the
connections, corresponding to local generators $D^\pm_{\pm a}$ and $T$ are
fixed constants. To obtain a general solution, one has to perform a
general local $\cH$-rotation of the constructed covariant derivatives.
Nevertheless, this particular solution is convenient for the component
analysis of the superfield models considered in the following sections.

{\bf 3.3 $\Os$-covariant derivatives in $M^{7|8}$.}

Here we study $\Os/U(2)$ derivatives and
their relation to the derivatives constructed in the previous section.
We denote the set of these derivatives by the same letters as the
corresponding generators of $\os$ superalgebra :
\bea
\mbox{local rotations} && S_\as,\; T, D^\pm_{\pm a} \cr
\mbox{covariant derivatives} && D_I = \{ \S_{\pm\pm\as},\; D^\pm_{\mp a},\;
D \}  \; \mbox{and}\; D_\a = \{ D^\pm_{\pm a} \}.
\label{d17-5}
\eea

First, let us consider the coset superspace $\cH/U(2)$, parametrized by
four odd coordinates $\q$. Let us call $D_{\pm a}$ the corresponding odd
covariant derivatives with local rotations $S_\as$, $T$. The algebra of
covariant derivatives coincides with
the (real form of)  $A(1|0)$ superalgebra
written out in Eq.(\ref{a21}). Note that every superfield $\Phi$ in this
supermanifold is described by $2^4 = 16$ coefficients of the decomposition
by power series in odd coordinates $\q$. Since the covariant derivatives
$D_{\pm a}$ are independent the following projections can
be chosen for these coefficients:
\bea
&\F|, \quad D_{\pm a} \F|, \quad
D^a_\pm D_{\pm a} \F| \equiv (D_\pm)^2 \F|, \quad
[D_{+a}, D_{-b}] \F|,&   \cr
&D_{\mp b} (D_\pm)^2 \F|, \quad
(D_\mp)^2 (D_\pm)^2 \F|&
\label{d18}\eea
where $\F|$ denotes the null component $\F(\q=0)$.

Now let us discuss the
relation of $\Os/U(2)$ covariant derivatives in $M^{7|8}$ to the
odd derivatives for $\cH/SU(2)$ and $\MM$ derivatives
constructed in the previous subsections. Here it is convenient to introduce
a shorthand notations. $\check D_I = \{\check\S_{\pm\pm\as},\; \check D_{\mp
a}^{\pm},\; \check D\}$ are covariant derivatives from
(\ref{d13},\ref{d16},\ref{d17});
$l_\a =\{D_{\pm a}\}$ are
just introduced odd derivatives in $\cH/U(2)$, while $\cL_\a =
\{D^\pm_{\pm a}\}$ are odd local rotations from (\ref{d16});
$S_a = \{S_\as, T\}$ are bosonic generators of local algebra $U(2)$;
at last $x=\{z,r\}$ are
coordinates  of the superspace $\bar M^{7|4}$. Then the derivatives $\check D_I$ and
$l_\a$ have the following structure:
\bea
\check D_I &=& e_I + {\o_I}^\a (x) \cL_\a + {\o_I}^a(x) S_a
\label{d19}\\
l_\a &=& e_\a + {\o_\a}^a(\q) S_a
\label{d19-5}
\eea
where $e$'s are vielbein vector superfields and $\o$'s are
connections\footnote{It follows from (\ref{d16}) that ${\o_I}^\a$
are constants in that particular basis for $\MM$ derivatives.}.
The derivatives $\check D_I$ satisfy the superalgebra
$\os$, while $l_\a$ form the $A(1|0)$ superalgebra
\bea
[\check D_I,\check D_J] &=& f^K_{IJ}\check D_K + f^\a_{IJ}\cL_\a +
f^a_{IJ}S_a,\label{d20}\\
{}[l_\a, l_\b] &=& f^\g_{\a\b} l_\g + f^a_{\a\b} S_a
\label{d21}\eea
where the structure constants $f$'s can be found in
(\ref{110}--\ref{1-10},\ref{13c}--\ref{18c}). Let us seek for
$\Os/U(2)$ covariant derivatives $D_I$, $D_\a$ in the following
form:
\bea
D_I &=& {E_I}^J(x,\q) e_J + {E_I}^\a(x,\q) l_\a +
{\O_I}^a(x,\q) S_a,\label{d22}\\
D_\a &=& l_\a, \label{d22-1}\\
{E_I}^J(x,0) &=& \d_I^J, \qquad {E_I}^\a(x,0) = {\o_I}^\a, \qquad
{E_I}^a(x,0) = {\o_I}^a.
\label{d23}\eea
Eqs.(\ref{d23})
give the initial conditions for the vielbeins $E_I$ and the connection
$\O_I$. Then we have to impose the following equations
which define the dependence of $D_I$ on $\q$'s:
\be
[l_\a, D_I] = f^J_{\a I} D_J.
\label{d24}\ee
Calculating the supercommutator and using the relations $[e_\a, \check D_I]
= 0$ and $[l_\a, S_a] = f^\b_{\a a} l_\b$, we obtain differential
equations for $E_I$, $\O_I$ of the following form:
\be
e_\a {E_I}^J(x,\q) + \ldots = 0, \qquad e_\a {E_I}^\b(x,\q) + \ldots = 0,
\qquad e_\a {\O_I}^a(x,\q) + \ldots = 0
\label{d25}\ee
with dots standing for terms without derivatives. This system of
differential equations is self-consistent
as it is checked using Jacobi identities for the
constants $f$, Eqs.(\ref{d22-1}) and the identity
$$
[l_\a,[l_\b,D_I]] + [l_\b,[l_\a,D_I]] + [[l_\a,l_\b],D_I] = 0.
$$
The self-consistency of the equations (\ref{d25})
and the linear independence of
the vector fields $e_\a$ assures the existence and uniqueness of a
solution for $E_I$ and $\O_I$ with the initial conditions (\ref{d23}).
To ensure
that this solution gives $\MM$ derivatives we have to check the
supercommutator $[D_I, D_J]$.
Due to linear independence of operators $D_I$, $l_\a$ and $S_a$ in every
point, the supercommutator $[D_I,D_J]$ can be expanded
in this basis, with some function coefficients $f(x,\q)$:
\be
[D_I, D_J] = f^K_{IJ}(x,\q) D_K + f^\a_{IJ}(x,\q) l_\a +
f^a_{IJ}(x,\q) S_a
\label{d26}\ee
Then $f(x,0) = f$ -- the structure
constants of the superalgebra $\os$. Really, the operator $l_\a$ in
Eq.(\ref{d26}) (and also inside operators $D_I$ (\ref{d22}))
can be replaced by the local generators
$\cL_\a$ with the same algebra and action on $D_I$ (\ref{d24}).
The null component
of the obtained operator $D_I^{l\ra\cL}$ is equal to $\check D_I$,
whence the null component of the equation (\ref{d26})
converts into (\ref{d20}) that provides the equation
$f(x,0) = f = {\rm const}$.
Then, taking the commutator of $l_a$ with the
equation (\ref{d26}), we get the differential equations on $\q$-dependence of
$f(x,\q)$. With the use of (\ref{d21},\ref{d24})
these equations can be reduced to form
\be
e_a f(x,\q) + \ldots = 0,
\label{d27}\ee
the dots are standing for terms without derivatives. These equations have the
unique solution which is constant, $f(x,\q) = f$. So we have proved
that the new covariant derivatives $D_I = \{\S_{\pm\pm\as},\; D^\pm_{\mp a},\;
D\}$ and $D_\a = \{D^\pm_{\pm a}\}$ satisfy the superalgebra $\os$.

A general superfield in $\Os/U(2)$  is represented locally by a set of
functions $\Phi_{p a(k)}$, carrying a finite-dimensional representation of
$u(2)$ algebra with $a(k)$ representing basis of $su(2)$ rank-$k$
symmetric spinor representation, and $p/2$ being the $u(1)$ $T$-weight.
In this paper only the case $k=0$ appears.

Analogously to the superfields in superspace $\cH/U(2)$ (\ref{d18}), all the
content of an arbitrary superfield $\F_p(z,r,\q)$ in superspace $M^{7|8}$
is given by the following component superfields in $\bar M^{7|4}$:
\bea
&\F_p|, \quad D^\pm_{\pm a} \F_p|, \quad
D^{\pm a}_\pm D^\pm_{\pm a} \F_p| \equiv (D^\pm_\pm)^2 \F_p|, \quad
[D_{+a}, D_{-b}] \F_p|,&   \cr
&D^\mp_{\mp b} (D^\pm_\pm)^2 \F_p|, \quad
(D^\mp_\mp)^2 (D^\pm_\pm)^2 \F_p|.&
\label{d28}\eea
The initial conditions (\ref{d23}), together with
Eqs.(\ref{d16},\ref{d17}) and the fact that
${\o_I}^\a$ in (\ref{d22}) are constant superfields gives us the following
component rules:
\bea
(\S_{\pm\pm\as} \F_p)| &=& \check\S_{\pm\pm\as}(\F_p|), \label{d29}\\
(D^\pm_{\mp a} \F_p)| &=& \pm\sqrt{2}\check\S_{\mp a} (\F_p|) +
(D^\mp_{\mp a} \F_p)|, \label{d30}\\
(D \F_p)| &=& -\hf \check\S(\F_p|) + \hf (T\F_p)|
\label{d31}\eea
where in (\ref{d30}) the second
term in r.h.s. is another independent component of
the superfield $\F_p$ while $T\F_p \equiv p/2 \,\F_p$ in the second term of (\ref{d31})
is the local rotation.

\sect{The chiral superfields on $M^{7|8}$. \label{s4} }

We introduce the main building blocks of the $N=2$ covariant equations to be
constructed, the {\it strongly chiral} fields on $M^{7|8}$.

A one-component superfield on $M^{7|8}$ is called {\it weakly chiral} if
it satisfies the constraint
\be \label{41}
E_a \F_p = 0 ,
\ee
where $E_a$ is one piece from the four odd covariant derivatives pieces on
$M^{7|8}:$
\be \label{e}
E_a = D^{+}_{+a} \;{\mbox {or}}\;  D^{+}_{-a} \;{\mbox {or}}\;
D^{-}_{+a} \;{\mbox {or}}\; D^{-}_{-a}.
\ee

The {\it strongly chiral} fields are defined to be weakly chiral w.r.t.
two special pieces from Eq.(\ref{e}). Namely, the
{\it strongly chiral field of a first kind} is defined by the
constraints
\be \label{c1}
D^{+}_{+a} \g_p = D^{-}_{-a} \g_p =0
\Rightarrow p=0.
\ee
It is clear from the commutation relations (\ref{110ad}) that in this case a
strongly chiral field should carry zero $T$-weight.

The {\it strongly chiral field of a second kind} is defined by the
constraints in which  the two covariant derivatives pieces have
equal $\G$-weights and, therefore, commute:
\be \label{42}
D^{+}_{+a} \F_p = D^{+}_{-a} \F_p =0
\ee
or
\be \label{42'}
D^{-}_{+a} \F_p = D^{-}_{-a} \F_p =0
\ee
In these cases, there are no restrictions on the $T$-weight.
Note that the superfield $\g$ can be real, $\g=\g^\da$, while
$\Phi_p$ is necessarily complex,
as the complex congugation $"\dagger "$
maps the covariant derivatives
$D^{+}_{+a} , D^{+}_{-a}  $ to
$D^{-}_{-a} , D^{-}_{+a}$ (Eq. (\ref{14'}).

Let's study the $M^{7|4}$ component content of
strongly chiral fields along the
previous section lines. It is clear that
the strongly chiral field of the first
kind possess the unique {\it unconstrained} component $\g(z,r) =\g_0 |$, the
rest components like $(D^{+}_{+a} \g_0) |$ are zero by virtue of constraints
(\ref{c1}).

Technically, the situation for the chiral fields of a second kind
is more involved.  Consider the constraint (\ref{42}) and introduce
the $M^{7|4}$ components of the field $\F_p$:
\be \label{43}
a=\F_p| \;,\; \j_a =( D^{-}_{-a} \F_p
)|\;,\; b= (( D^{-}_{-})^2 \F_p)|.
\ee
The rest components like $( D^{+}_{+a}
\F_p)|$ etc. are zero by virtue of the first
constraint from (\ref{42}).  We have
to analize how the remaining constraint $Y^{+}_{-a} \equiv D^{+}_{-a} \F_p =0 $
looks like in terms of the $\, a, \j_a , b \,$ fileds by studying the
components of this equation.  There are four groups of components
(since  $D^{+}_{+b} Y^{+}_{-a} =0$):
\bea
I: &
Y^{+}_{-a} | &= \j_a + \sqrt{2} \chs_{-a} a =  0 \\
II: &
D^-_{-a} Y^+_{-a} | &= \chs_{--a(2)} a - \sqrt{2} \chs_{-a} \j_a =0 \\
III: &
D^{-a}_{-} Y^{+}_{-a} | &= b + \sqrt{2} \chs^a_- \j_a  = 0 \label{44} \\
IV:&
(( D^{-}_{-})^2 Y^+_{-a})|  &= 2\chs_{--a(2)} \j^a +
\sqrt{2} \chs_{-a} b = 0.
\eea
Here we have used formulas (\ref{d29}--\ref{d31}) and $\Osp$
covariant derivatives $\chs$ in supermanifold $\bar M^{7|4}$.
The first and the third groups show that $\j_a$ and $b$ are expressed via
the unconstrained field $a$.
The second and the fourth groups prove to be the consequences of the
first and the third ones and do not lead to any restriction on $a$. Thus,
the chiral field of the second kind $\F_p$ is equivalent to the {\it
unconstrained} $M^{7|4}$-superfield, being expressed by the first projection
$a=\F_p|$.

As far as the component
content of either of the considered strongly chiral
superfields reduces to one {\it unconstrained} field in $\bar M^{7|4}$,
they can be chosen on the role of dynamical variables in
physical models. Though both these components
are similar superfields in $\bar M^{7|4}$, their
$N=2$ supersymmetry transformations  differ.
One can show that for $\g=\g_0|$ the transformations of the second
supersymmetry (\ref{A15}), parametrized by Killing scalar $t$,  are
realized by the operator $\cM$ (\ref{A24}). This means that
the $osp(2|4)$
representation carried by $\g$ is an ordinary regular
representation on the functions in
homogeneous space. However, for $a=\F_p|$, the second supersymmetry is given
by the operator $\cN$ whose action on the real and imaginary parts of $a$
is similar to the action on the superfields $Z$ and $\r$ in (\ref{A37}).

The {\it strongly chiral fields of a third kind } are defined to obey the
constraints
\be
D^+_{-a} \F_p =D^-_{+a} \F_p = 0 \Rightarrow {\tilde T}\F_p \equiv ( 3T -4D )
\F_p =0
\ee
One may show that such a field possesses just one constrained component on
$\bar M^{7|4}$: $\f =\F_p |$,
$(3p+4\chs)\f =0$,
and all higher components are expressed via $\f$.

\sect{Equations of motion for higher spin superfields with manifest
$N=2$ supersymmetry.\label{s5}}

  In this section, we present explicitly $N=2$ supersymmetric equations
equivalent to (\ref{A361}--\ref{A363}). They are the dynamical equations
for higher spin fields formulated in superspace $M^{7|8}=\Os/U(2)$.
According to App. A, in this superspace there exist the covariant
derivatives $\S_{\pm\pm\as}$, $D_{\pm a}^{\pm}$, $D_{\mp a}^{\pm}$ and $D$
that satisfy the $\os$ superalgebra
(\ref{110}--\ref{1-10},\ref{13c}--\ref{18c}) together with local rotations
$S_\as$ and $T$.  The explicit expressions for these derivatives in terms
of $\Osp$-covariant objects are derived in Sect. 3, and used below in the
component analysis.

Let us begin to formulate the theory.
The superfields on $M^{7|8}$ are characterized by their $su(2)$ indices
and $T$-weights. Dynamical variables are strongly
chiral superfields on $M^{7|8}$:
real $\g = \g^\da$, complex $\F_6 \equiv \A$ and its conjugated
$\F_{-6} \equiv \B = (\A)^\da$ subjected to the constraints
\bea
D^+_{+a} \g = D^-_{-a} \g = 0, && T \g \equiv 0, \cr
D^+_{-a} \A = D^+_{+a} \A = 0, && T \A \equiv 3 \A, \label{501}\\
D^-_{+a} \B = D^-_{-a} \B = 0, && T \B \equiv -3 \B. \non
\eea
Equivalently, $\g$ is a strongly chiral superfield of a first kind and
$\A$ is of a second kind one (see Sec. \ref{s4}). The r.h.s. of the last
equations are not actual constraints in $M^{7|8}$, they just exhibit the
$T$-weights  of the superfields introduced.
The peculiar notation $\A$ for the $\Phi_6$ superfield will be justified
in a special on-shell gauge (\ref{512}) $\g = 0$, where $\G \A = 2 \A,
\, S  \A = - 2\A$, thus this field has definite weights w.r.t. both
$\G$ and $S$ generators like generators $D^{\pm}_{\pm a}$ etc. do,
see Eqs.(\ref{15c},\ref{16c}).

The higher-spin dynamics is described by two (one real and one complex)
equations of motion for these superfields:
\bea
&&D_+^{-a} D^-_{+a} \A + D_-^{+a} D^+_{-a} \B =
8i ([D_-^{+a}, D^-_{+a}]_{-} - 2) \g,   \label{502}\\
&&i(D-2)\A = D_-^{+a} D^+_{-a} \g,
\label{503}
\eea
These equations have gauge symmetry with real first kind strongly chiral gauge
parameter $\ve$,
\bea
\d \A &=& D_-^{+a} D^+_{-a} \ve, \qquad
\d \B = - D_+^{-a} D^-_{+a} \ve, \label{504}\\
\d \g &=& iD \ve, \qquad D^\da = - D,  \label{505}\\
D^+_{+a} \ve &=& D^-_{-a} \ve = 0, \qquad T \ve \equiv 0 \non
\eea
The invariance of the second equation follows from $[D, D_-^{+a}] =
D_-^{+a}$ and of the first one -- from the relation
\be
[D_+^{-a} D^-_{+a}, D_-^{+a} D^+_{-a}] \ve
= -8 ([D_-^{+a}, D^-_{+a}] - 2) D \ve,
\label{506}
\ee
which is valid for scalars with zero weight w.r.t. generator $T$.
The null $\bar M^{7|4}$ component of $\ve$ is nothing but
the gauge parameter of GKS formulation (\ref{A14}).

    Using the previous section results one concludes that the component
content of superfields (\ref{501}) is characterized by one real and one
complex scalar in $\bar M^{7|4}$. This means that neither auxiliary nor pure
gauge degrees of freedom were added to the dynamical content of the model
(\ref{A30}). It can be considered as the reflection of the fact that
though the $N=2$ invariance (\ref{A37}) of the equations of motion
(\ref{A361}--\ref{A363}) is implicit, its algebra closes off-shell and
without gauge additions (\ref{A35}). One can see that the manifest $N=2$
supersymmetry (realized on superfields (\ref{501}) in terms of a canonical
left action of a group on its homogeneous space), gives the laws
(\ref{A37}) on the component level. This canonical action in $M^{7|8}$ is
given by a Killing operator analogous to operator $\cK$ (\ref{A23}) in
$\bar M^{7|4}$, and its component form can be derived with help of rules of
component analysis formulated in the end of subsection 3.3.
It agrees with the fact that the components $Z$, $\r$
and $\g$ from (\ref{A30}) decompose in two $N=2$ superfields $\A$ and
$\g$ (\ref{501}). The coincidence of this explicit
symmetry with (\ref{A37}) means that it does not leave the action
invariant (see Sect.2). That is why though the proposed equations of motion are
Lagrangian and the corresponding action is $N=2$-supersymmetric, it cannot
be expressed in explicitly $\Os$-invariant form via the superfields $\A$ and
$\g$.  Nevertheless, there still remains the question if there exists a
formulation with an explicit realization of the supersymmetry $\tilde\d_t$
(\ref{A32}). In such a formulation, the action would have to possess the
manifestly invariant form.

    Now we investigate  the component form of the equations
(\ref{502},\ref{503}).  To begin with, consider
the null components of these equations. It was shown in
Sect.4 that the superfield $\g$ contains only one
independent real component (which we
also denote $\g$) and the superfield $\A$ has only one
independent complex component
$a$.
\bea
&&\g(z,r) = \g(z,r,\q)| \equiv \g(z,r,\q=0), \label{507}\\
&&a = \A|, \qquad D^-_{-a} \A| \equiv \j_a =
-\sqrt{2} \check\S_{-a} a, \label{508}\\
&&(D_-^-)^2 \A| \equiv b = -\sqrt{2} \check\S_-^a \j_a, \non
\eea
with $\check\S_{-a}$ defined in (\ref{d14}).

Now we are in a position to consider $\q=0$
components of the equations (\ref{502},\ref{503}).  To calculate them one
should ,
first of all,
pass to $\Osp$-covariant derivatives $\check\S_I$ by the formulas
(\ref{d29}--\ref{d31}) and definitions (\ref{507},\ref{508}) for components.
Second, one should use the expressions (\ref{d9}--\ref{d14}) for these
derivatives. At last, one has to pass to $SL(2,C)$ covariant indices by
(\ref{d7}). Then all the dependence on the small vielbein and its derivatives
(\ref{d8-1}--\ref{d8-5}) disappear and one is rest with ordinary $N=1$ AdS superspace covariant equations
on $\bar M^{7|4}$. In so doing we come to the
following two equations:
\bea
&&-(i(\cP-\cPB) + \hf (\cQ+\cQB)) a - c.c. =
16i (\cP + \cPB -1) \g, \label{509}\\
&&-\frac i2 (\na+1) a = \{i(\cP-\cPB) - \hf (\cQ+\cQB)\} \g. \label{510}
\eea
These equations convert into (\ref{A361}--\ref{A363}) after
the identification $a = -4(Z+i\r)$. The real equation (\ref{509})
turns into
(\ref{A363}) while the complex equation (\ref{510}) gives the sum
(\ref{A361}) + $i$(\ref{A362}). So, it is proved that the
dynamical equations (\ref{502},\ref{503}) for superfields (\ref{501})
lead to the original higher spin GKS equations (\ref{A361}--\ref{A363}).

Let us discuss the higher components of equations of motion
(\ref{502},\ref{503}) and prove
that they do not imply any new information. First note that the second
equation (\ref{503}) is strongly chiral, it satisfies identically the same
constraints the superfield $\A$ does. A strongly chiral superfield of this type
has only one independent component -- its null component (see the previous
section). Hence the equation (\ref{510}) in superspace $\bar M^{7|4}$
is equivalent to the superfield equation (\ref{503}).

The situation with the first
equation (\ref{502}) is a bit more complicated.
It is convenient to consider this
equation in the gauge $\g = 0$ which 
admissibility is shown after Eq.(\ref{A31}).
Let us introduce a weakly
chiral superfield $B = D_+^{-a} D^-_{+a} \A$, $D^-_{+a} B \equiv 0$. Then, the
equations of motion (\ref{502},\ref{503}) in this gauge take the form
\bea
E \equiv B+B^\da &=& 0, \label{511}\\
(D-2) \A &=& 0
\label{512}\eea
To ascertain that all the components of the equation (\ref{511})
are derivatives of the null one we use Eq.(\ref{512}) which is
already shown to
be equivalent to (\ref{510}) with $\g=0$.
Thereby (\ref{502}) follows from the couple
of equations (\ref{509},\ref{510}), not from the only Eq.(\ref{509}).
Using the algebra (\ref{110}--\ref{1-10},\ref{13c}--\ref{18c}) of covariant
derivatives the following relation can be derived straightforwardly
$$
D^+_{-a} B = D^+_{-a} D_+^{-b} D^-_{+b} \A = 8 D^+_{-a} (2-D) \A
$$
Thus we come to the additional
constraint $D^+_{-a} B = 0$
on superfield $B$ (note that $D^-_{+a} B =
D^+_{-a} B = 0$ implies $D B =0$ as it follows from (\ref{1-10})).
So, if Eq.(\ref{512}) is satisfied, the superfield $E$ obeys the constraints
\be
D^-_{+a} E = D^+_{-a} E = 0
\label{513}\ee
Therefore, $E$ is a third kind strongly chiral superfield which possesses only
one independent component $E|$ (see Sec. \ref{s4}), this completes
the proof.

\sect{Picture changing and manifestly $N=2$ supersymmetric equations on
$M^{7|4}$.\label{s6}}

In this section, we show that the whole system of equations and
constraints on $M^{7|8}$ (\ref{501}--\ref{503})
may be equivalently reformulated as a system
of equations and constraints on $M^{7|4}= Osp(2|4)/\cH$. This
framework seems to be quite natural since, from the $N=1$ viewpoint, $M^{7|4}$
is just that manifolfd the original manifestly $N=1$ supersymmetric theory was
defined.

Calculations of this section do not deal with  $N=1$ component expressions,
and $N=2$ covariance is kept manifest. It is convenient therefore to introduce
a universal notation: given an $osp(2|4)$ coset space, both the covariant
derivatives and local rotations are denoted by the same letter $D$
(except for local rotations $S_\as$ and $T$)
differing by the indices they carry according to their place in
$osp(2|4)$ superalgebra.
In what follows, we use two equivalent
notations for components of $\Os$ generators
$D_{\e\z}$, $D_{\eb\zb}$ and $D_{\eb\z}$. First, we use
the index notation introduced in App. C: $\e=(a,z)$, $\eb = (\bar
a,\bar z)$, etc., where $a$, $\bar a$ are $su(2)$-indices and $z$,
$\bar z$ denote the single bosonic value of superindices $\e$, $\eb$
respectively. Then, we also use the notation explicitly indicating
$T$-weights. The following expressions include all these notations
\be \label{1117}
D_{\eb \z} =  \left( \ba{cc}
                                D_{\bar a b}    & D_{\bar a z}\\
                                D_{\bar z b}    & D_{\bar z z}
                                \ea \right) =
                \left( \ba{cc}  S_{ab}-\ve_{ab} T & -D_{+a}\\
                                D_{-b}            & 2T
                                \ea \right),\quad
  C^{\z \eb}  D_{\eb \z} =0 ,
\ee
\be \label{1118}
D_{\e \z} =  \left( \ba{cc}
                               D_{ab} & D_{az}\\
                               D_{zb} & D_{zz}
                                \ea \right)=
               \left( \ba{cc}
                               D_{2ab} & -D_{3a}\\
                                D_{3b}         & 0
                                \ea \right),\quad
   D_{\e \z} =-(-)^{\e \z} D_{\z \e}    ,
\ee
\be \label{1119}
D_{\eb \zb} =  \left( \ba{cc}
                               D_{\bar a\bar b} & D_{\bar a\bar z}\\
                               D_{\bar z\bar b} & D_{\bar z\bar z}
                                \ea \right)=
               \left( \ba{cc}
                               D_{-2ab} & -D_{-3a}\\
                               D_{-3b}  & 0
                                \ea \right),\quad
   D_{\eb \zb} =-(-)^{\eb \zb} D_{\zb \eb}    ,
\ee
The rest piece of $osp(2|4)$, the scalar derivative is always denoted $D$.
Similar notations are used
for superfields, for example, a supersymmetric tensor superfield
$\f_{\e_1\e_2\e_3}$ contains three $SU(2)$-irreducible component fields
(\ref{663}). At last, the covariant derivatives in superspaces
$\M8$ and $M^{7|4}$ are denoted identically; the manifold, on which
a derivative lives, is specified by the superfield this
derivative acts.

To start with we observe that there exist special constrained superfields
on $M^{7|8}$  which are equivalent to the unconstrained $M^{7|4}$ ones.
Indeed, consider a superfield $\Phi^{\bf A} (z,r,\q_{\pm a})$ on $M^{7|8}$, which
takes its values in an $su(2,0|1,0)$ fdr
(or in $A(1|0)$ fdr, which is the complexification of $su(2,0|1,0)$ )
, the index ${\bf A}$ just labels this fdr's
basis.  Let ${\bf D_i}$ be $su(2,0|1,0)$ generators,
${\bf {\hat D}_i}$ be their image in the covariant derivatives algebra
on $M^{7|8}$ and  ${\bf {(D_i)^A}_B}$ be their form in the $su(2,0|1,0)$ fdr.
Impose the constraints
\be \label{61}
{\bf {\hat D}_i} \Phi^{\bf A} = {\bf {(D_i)^A}_B} \Phi^{\bf B}.
\ee
The even part of these constraints is satisfied identically as for
$S_{ab}, T$ generators the equations (\ref{61}) are just the definition of
how these generators enter the whole covariant derivatives algebra (see App.
A). The constraints may be solved in a standard way, (at least locally on
$\M8$) in terms of local fields on $M^{7|4}$ which is identified with the
subspace $\q=0$ (see Sect. \ref{s3}).  A general solution reads
\be  \label{62}
{\F}^{\bA} (z,r, \q_{\pm a}) = exp (\q^{+a} {\bD}_{+a} +
\q^{-a} {\bD}_{-a}){^{\bA}}_{\bB} {\F}^{\bB} (z,r, 0)
\ee
It establishes a bijection between the $\M8$ constrained fields of
the form (\ref{61}-\ref{62}) and the unconstrained local tensor fields ${\F}^{\bB}
(z,r, 0) \equiv \f^{\bB} (z,r)$ on $M^{7|4}$ transforming in the same
$su(2,0|1,0)$ fdr as the original $\M8$ ones.
If the representation labeled by $\bf A$ is irreducible, then, for
every fixed value of $\bf A$, the component fields
${\F}^{\bA}|\,,D_{\pm a} {\F}^{\bA}|
\ldots $ are obviously in one-to one correspondence with
the set of all $\f^{\bB} (z,r)$ components;
the pictures arising for different $\bf A$ are connected
by an $A(1|0)$ transformation.

In order not to be too abstract it is pertinent to introduce now an  example
and then illustrate all the ideas on it, the generalization to an
arbitrary fdr is straightforward.
Consider the $A(1|0)-$representation $(2|2)_p, p\neq \pm1$ (see Appendix C,
Eqs. (\ref{a210}-\ref{a211})\footnote{In this section, we use extensievely the
results and notation of Appendix C}.), so
the field's $\Phi_{\bf A}$ components are ($p, p \pm 1$ labels the $T$-weight
and $a$ is $su(2)$ spinor index)
$$
\Phi_{\bf A} = ( \F_{pa}\;,\; \F_{p-1}\;,\; \F_{p+1} ),
$$
while the constraints (\ref{61}) read (compare to (\ref{a211}))
\be \label{63}
\ba{cc}
D_{+b} \left( \ba{c} \F_{pa} \\ \F_{p+1} \\ \F_{p-1} \ea \right) =
\left( \ba{c} \ve_{ba} \F_{p+1} \\ 0 \\ \frac{1-p}{2} \F_{pb} \ea \right) ; &
D_{-b} \left( \ba{c} \F_{pa} \\ \F_{p+1} \\ \F_{p-1} \ea \right) =
\left( \ba{c} \ve_{ba} \F_{p-1} \\ \frac{1+p}{2} \F_{pb}\\ 0 \ea
\right).
\ea
\ee
We see that all
the content of superfields $\Phi_{\bf A}$ reduces to the component
fields of superfield $\Phi_{p+1}$ because
the part of relations (\ref{63}) leads to
\be  \label{63'}
\F_{pa} =2 (1+p)^{-1} D_{-a} \F_{p+1} \;\;;\;\; \F_{p-1} = (D_{-})^2 \F_{p+1},
\ee
while the rest constraints imply
\be \label{63'a}
D_{+a} \F_{p+1} =0.
\ee
So,
one can equivalently describe any
$\cH$-tensor
superfield
on $M^{7|4}$ in two ways: first, as a cross section of the
$\cH$-bundle
on $M^{7|4}$,
second, as a constrained superfield on $M^{7|8}$.  We shall refer to the first
way as to $M^{7|4}$-picture and to the second one as to $M^{7|8}$-picture.  For
instance, it is seen that {\it the weakly chiral (w.r.t. $D_{+a}-$ derivative)
one-component superfield $\F_{p+1}$ on $M^{7|8}$ is equivalent to the
$\f_{i_p}$ unconstrained superfield on $M^{7|4}$ $, p \neq \pm 1$, or they form
the two pictures of the same $osp{(2|4)}$ superalgebra representation}.

Recall that the $N=2$ covariant derivatives on $M^{7|8}$ constitute, along
with $su(2,0|1,0)$-inner rotations, the full $osp(2|4)$-superalgebra which
decomposes into direct sum of four graded subspaces w.r.t. $su(2,0|1,0)$
(\ref{114}--\ref{116}).
Therefore, for the covariant derivative set the constraints
(\ref{61}) are satisfied identically.

The main result of this section
is the observation that all equations and constraints of  Sect. \ref{s5}
constitute the $M^{7|8}$ -picture of some $M^{7|4}$ tensor equations.
First note that, as is obvious from the above analysis, the constrained
fields  $\F_6 \equiv \A$ and $\g$ (\ref{501})
may be viewed as $M^{7|8}$ -picture of
$M^{7|4}$ tensor fields $\f_{i_5}$ and $\g$, (the strongly chiral field $\g$
corresponds to the trivial $(1|0)$-representation). Recalling the formula
(\ref{a216}) from the App. C we
represent the $\f_{i_5}$ field as the supersymmetrized product of
three $\x$ fdrs : $\f_{\x(3)}$, with the highest weight component being
identified with $\f_6$ :
\be \label{663}
\F_6| = \f_6 = \f_{z z z}  \;,\; \F_{5a}| = \f_{5a} =\f_{a z z}\;,\;
\F_4| = \f_4  =  \varepsilon^{ba} \f_{ab  z} \ee
The  $\F_6, \F_{5a}, \F_{4}$  supefields are assumed to obey  the equations
(\ref{63}--\ref{63'a}) with $p=5$. It is clear from the above discussion
that, in $M^{7|4}$  picture, the
counterparts of constraints $D_{+a} \Phi_6 =0$
are absent. At the same time, the constraint
\be \label{67}
D_{3a} \F_6 =0
\ee
turns into a tensor equation
\be \label{68}
D_{\x_1(\x_2} \f_{\x_3 \x_4 \x_5 )} = 0.
\ee
The round brackets denote supersymmetrization, as usual. As covariant
derivatives  $D_{\x_1\x_2}$ are superantisymmetric in its indices, the last tensor
equation is characterized by the Young tableaux of the form

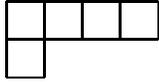
\unitlength=0.63mm \special{em:linewidth 0.4pt} \linethickness{0.4pt}
\begin{picture}(30.00,40.00)(-50.0,117.0)
\put(10.00,150.00){\line(1,0){32.00}}
\put(42.00,150.00){\line(0,-1){8.00}}
\put(42.00,142.00){\line(-1,0){32.00}}
\put(10.00,142.00){\line(0,1){8.00}}
\put(10.00,142.00){\line(0,-1){8.00}}
\put(10.00,134.00){\line(1,0){8.00}}
\put(18.00,150.00){\line(0,-1){16.00}}
\put(26.00,150.00){\line(0,-1){8.00}}
\put(34.00,150.00){\line(0,-1){8.00}}
\end{picture}

To prove the relation (\ref{68}), it is sufficient to observe that
the first $M^{7|4}$ projection of (\ref{67}) presents the highest-weight
component of (\ref{68}) and the next projections,
associated with all $D_{\pm a}$ derivatives of (\ref{67}),
are identified with the rest components of the tensor equation (\ref{68}).
It is important that one should not break the $\cH$-covariant structures by
passing to the $N=1$ components as in Sections \ref{s3} and \ref{s4}. Indeed,
consider the basis in $\os$ superalgebra  introduced in
(\ref{d15-3},\ref{d15-5}). Then it follows from Eqs.(\ref{d29}--\ref{d31})
that for combinations $\S_I$ (\ref{d15-5}), the relation holds:
\be \label{69}
(\S_I \F^{\bf A})| = \check\S_I (\F^{\bf A}|)
\ee
Due to the constraints (\ref{61}) the same equation is valid for all
generators $D_{\eb\z} \in \os$:
\be \label{69'}
(D_{\eb\z} \F^{\bf A})| = D_{\eb\z} (\F^{\bf A}|)
\ee
where in r.h.s. $D_{\eb\z}$ acts as the local rotation on index $\bA$.
So, these equations are satisfied for every generator of $\os$ and
give the $\os$-covariant way to pass to $M^{7|4}$-components of superfields
in $\M8$.
For example, one has
\be \label{691}
(D_{3a} \F_6)| = D_{3a} (\F_6|) =  D_{z a} \f_{zzz} = 24 D_{z(a}
\f_{zzz)}
\ee
for the first $M^{7|4}$ projection, or
\be
\ba{l} \label{692}
(D_{-b} D_{3a} \F_6)| = \\ \\= D_{-b} D_{3a} (\F_6|) = [D_{-b} D_{3a}] \f_6
- D_{3a} D_{-a} \f_6 = \S_{++ba} \f_6 - 3 D_{3a}  \f_{5b} = \\ \\ =
D_{-b} D_{z a} \f_{zzz} =  D_{ba} \f_{zzz} - 3 D_{z a}
\f_{b  zz} = 24 D_{a(b} \f_{zzz)}
\ea
\ee
for the "$D_{-b}$"-projection, and so on. This proves the Eq. (\ref{68}).

To find the picture changed $M^{7|4}$
reformulation of two equations (\ref{502},\ref{503})
introduce supersymmetric third-rank
tensor operators (for the definition of invariant seven-rank
${\tilde C}$-tensors, see App.C, Eqs.(\ref{a224}-\ref{a228}))
\be \label{664}
\ba{c}
\D_{\e_1 \e_2 \e_3} = D^{\etab_2 \etab_1} D^{\xb_2 \xb_1} {\tilde C}_{\xb_1
\xb_2 ;} {}_{\etab_1 \etab_2 ;} {}_{\e_1 \e_2 \e_3} \\ \\
\D_{\eb_1 \eb_2 \eb_3} = D^{\eta_2 \eta_1} D^{\x_2 \x_1} {\tilde C}_{\x_1
\x_2;} {}_{\eta_1 \eta_2 ;} {}_{\eb_1 \eb_2 \eb_3} ,
\ea
\ee
possessing the components
\be \label{665}
\ba{cc} \label{612}
\D_{zzz} = -4(D_3)^2 & \D_{{\bar z}{\bar z}{\bar z}} = -4(D_{-3})^2 \\
\D_{zzc} = -8\left( D_3^a  D_{2ca} \right)  &
\D_{{\bar z}{\bar z}{\bar c}} = -8\left( D_{-3}^a  D_{-2ca} \right) \\
\D_{zcd} = -2 \ve_{cd} \left( D_2^{ab} D_{2ba}  \right)&
\D_{{\bar z}{\bar c}{\bar d}} = -2 \ve_{cd} \left( D_{-2}^{ab} D_{-2ba}  \right)
\ea
\ee
Also, we need another operator of the form
\be \label{667}
\D_{\e \eb} = {D_{\e}}^{\etab} D_{\etab \eb} + (-)^{\e \eb}
{D_{\eb}}^{\eta} D_{\eta \e}
\ee
with the components
\be \label{613}
\ba{c}
\D_{z\,{\bar z}} = -(D_3^a D_{-3a} +  D_{-3}^a D_{3a}) \equiv -[ D_3
D_{-3}]_{-}\\ \\
\D_{a\,{\bar z}} = - 2 {D_{2a}}^b D_{-3b}\\
\D_{z\,{\bar a}} = - 2 {D_{-2a}}^b D_{3b}\\ \\
\D_{a\,{\bar b}} = {D_{2a}}^c D_{-2cb} -{D_{-2a}}^c D_{2cb}
+D_{-3a} D_{3b} - D_{3a} D_{-3b}
\ea
\ee
Then, the $M^{7|4}$ picture of equations (\ref{502},\ref{503})
looks as follows
(the $"[=]"$ symbol means $"="$ with an account of a sign factor
(App. B)):
\be  \label{614}
\ba{c}
-\frac{1}{4} \D_{\e(3)} \g = i (D -2) \f_{\e(3)}\\    \\
\frac{1}{4} \left( \Delta_{\e(3)} \f_{\eb(3)} - (-)^{\e(3) \eb(3)}
\Delta_{\eb(3)} \f_{\e(3)} \right) \;\;[=]\;\; -8i \left( C_{\e \eb}  C_{\e
\eb}
\Delta_{\e \eb} -2 C_{\e \eb} C_{\e \eb} C_{\e \eb} \right) \g
\ea
\ee
Indeed, the $"zzz"$-component of the first equation and the
$"zzz,{\overline{zzz}}"$
component of the second equation are
\be \label{668}
\ba{c}
-\frac{1}{4} \D_{zzz} \g = i(D-2) \f_{zzz} \\ \\
\frac{1}{4}  \left( \D_{zzz} \f_{\overline{zzz}} - \D_{\overline{zzz}}
\f_{zzz} \right)= 8i (\D_{z\,{\bar z}} -2) \g
\ea
\ee
which, after accounting
the component form of the operators (\ref{612},\ref{613}) are seen to
coincide with $M^{7|4}$-projection of Eq.
(\ref{502},\ref{503}) upon the identification $\Phi_6 \equiv \Phi^{++}_{--},
\Phi_{-6} = - \Phi^{--}_{++} $.
It remains to check that the rest components of
Eq.(\ref{614}) are in one to one correspondence with all
components
of $M^{7|8}$ equations (\ref{502}, \ref{503}). We have performed this
check explicitly and have found they match indeed.

The $M^{7|4}$ -picture of gauge invariance reads
\be \label{615}
\ba{c}
\d \phi_{\e(3)} = -\frac{1}{4} \D_{\e(3)} \varepsilon \;,\;
\\ \\ \d \g = iD \varepsilon
\ea
\ee
where $\varepsilon$ is a real scalar field presenting the unique
$M^{7|4}$ projection of the strongly chiral field of a first kind  $\varepsilon$
on $M^{7|8}$.

This is easy to prove since
the $"zzz"$ component of the first equation gives null component of
the gauge transformation
law (\ref{504}) and the rest components coincide with $\q=0$ projections
of the $D_{\pm a}$-derivatives  of
this law; the only component of the second equation is identical to the
only null component of (\ref{505}).
On the other hand, one may check the gauge invariance in
$M^{7|4}$-picture directly, by using the covariant derivatives algebra
(\ref{120}--\ref{125}).

Thus, we have observed that the higher spin equations
(\ref{502},\ref{503}) are well
defined on $M^{7|4}$ and derived such a picture. It is characterized by two
$M^{7|4}$-superfields $\phi_{\e(3)}$ and $\g$ satisfying the equations
(\ref{68},\ref{614}),  which turn out to be gauge invariant w.r.t.
transformations (\ref{615}).

A few remarks are in order.
Recall that the fdrs $\x(3) \xb(3)$ and $\x(2) \xb(2)$ have equal dimension
(see App.C), therefore there is no information loss if one takes the
first trace of the second equation (\ref{614})
by an invariant second rank tensor $C^{\zb \e}$ (\ref{a26} - \ref{a28}):
\be \label{669}
\ba{c}
\frac{1}{4} \left( {\Delta_{\e(2)}}^{\zb} \f_{\zb\eb(2)} - (-)^{\e(2) \eb(2)}
{\Delta_{\eb(2)}}^{\z} \f_{\z\e(2)} \right) [=] \\ \\
{}[=] \;\; -8i \left( \frac{4}{9}\, C_{\e \eb}  \Delta_{\e \eb} +\frac{1}{9}\,
C_{\e \eb} C_{\e \eb} {\Delta^{\zb}}_{\zb} +\frac{1}{3}\, C_{\e \eb} C_{\e \eb}
\right) \g
\ea
\ee
Now we apply the analysis of $\x(2) \xb(2)$ representation
from  Appendix C (\ref{a231} - \ref{a241}) to the l.h.s.
and the r.h.s. of the last equation.

Obviously, the r.h.s $\in H_{trace} \subset H^{7|8} $ where $H^{7|8} $ is
a subspace of double-traceless tensors and
$H_{trace}$ is defined by (\ref{a236}).
Therefore, the l.h.s. obeys this condition either.
The relation l.h.s $\in H^{7|8}$ leads to the covariant $\g$-independent
consequence of (\ref{669}):
\be \label{670}
\ba{c}
\frac{1}{4} \left( \Delta^{\eb(3)} \f_{\eb(3)} - \Delta^{\e(3)} \f_{\e(3)}
\right) = 0
\ea
\ee
Further, $H_{trace}$
is a general solution of the constraints (\ref{a240'} , \ref{a241})
so one gets
\be \label{6155}
\ba{c}
0 \;\; [=]\;\; {{C_{\e(2);}}^{\zb(2)}{}_{;\eb}}{}^{\z(2)}
\left(l.h.s.\right)_{\z(2) \zb(2)} \\ \\  0 \;\; [=] \;\;
{{C_{\eb(2);}}^{\z(2)} {}_{;\e}}{}^{\zb(2)} \left(l.h.s.\right)_{\z(2)
\zb(2)} ,
\ea
\ee
i.e. two additional $\g$-independent consequences of Eq.
(\ref{669}).
Here 'l.h.s.' means left hand side of the Eq.(\ref{669}).

In the $M^{7|8}$-picture, these consequences are expressed via the constraint
\be \label{671}
(D_{\pm})^2  \left(r.h.s.\right) =  (D_{\pm})^2 \left(
8i (\D_{z{\bar z}} -2) \right) \g = 0,
\ee
Therefore, the l.h.s.
should obey these equations either:
\be \label{672}
(D_{\pm})^2 \;\frac{1}{4}  \left(
\D_{zzz} \F_{-6} - \D_{\overline{zzz}} \F_{-6}\right) = 0
\ee
The $M^{7|4}$-projections of the $D_{\pm a}$ - derivatives  of the last equation
give rise to tensor equations (\ref{6155}), which include only $\phi_{\e(3)}$
and not $\g$.  All these additional equations are gauge-invariant as the
equations (\ref{68}, \ref{614}) are.

\sect{Conclusion.}

Let  us summarize the results. We succeeded in constructing two manifestly
$N=2$ covariant formulations for the equations of motion of
GKS theory \cite{gks}, which describes free massless
fields of all superspins in $D=4$ AdS space in terms of a few scalar fileds
on $\bar M^{7|4}$.
The first formulation is achieved
through employing the constrained (strongly chiral) superfileds in $N=2$
superspace $M^{7|8} = Osp(2|4)/U(2)$ while the second one uses a smaller
superspace $M^{7|4} = Osp(2|4)/SU(2,0|1,0)$ which, from $N=1$
veiwpoint, is just the same manifold the original GKS theory is defined.
The obtained equations have a neat form and, in a sense, look simpler than
$N=1$ ones.

Unfortunately, all of this does not give a possibility to construct an $N=2$
manifestly invariant form of the action, since the laws of global $N=2$
transformations, which follow from the superfield formalism developed
in the paper, differ from those leaving the action invariant by
terms proportional to the equation of motion multiplied by a {\it symmetric}
matrix.

However, there still exist a lot of possibilities to improve the situation.
Here we discuss how one can try to construct an explicitly $N=2$ invariant
action.
As far as
the action is not invariant w.r.t. transformations $\bar
\d$ (\ref{A37}),
one should make explicit the
supersymmetry $\tilde\d$ (\ref{A32}). These transformations have the structure of
a semidirect sum and an invariant subspace extracted by the
equation $\g=0$ exists.
Note that the algebra of global transformations
$\tilde\d|_{\g=0}$ is determined by the algebra of vector fields $\cN$
(\ref{A33}); the addition of scalar $t$ with an arbitrary factor does not change
the algebra. Thus there is a one-parameter set of multiplets, each multiplet
consists of two real superfields $(Z,\r)$ transforming through each other by
the laws differing from $\tilde\d|_{\g=0}$ by the addition of scalar $t$ with an
arbitrary factor.

On the other hand, it is possible to construct different versions of $M^{7|8}$
as $M^{7|8}_{(a,b)} = Osp(2|4)/SU(2)\oplus U(1)_{(a,b)}$ where the $su(2)$
generators are the same as for $M^{7|8}$ while the $U(1)_{(a,b)}$ generator is
taken as $a\G + b S$ (in the present paper we dealt with  $a=1,
b=-\frac{1}{2}$).  Then one has to study diverse chiral superfields on these
manifolds just analogously to the procedure of Sec. (\ref{s4}) and find those
reducing to unconstrained supefields on $\bar M^{7|4}$. The existence of the
superfields of such a kind is hinted by the observed evidence of one-parametric
family of nonequivalent $Osp(2|4)$ multiplets. It is probably that, at some
specific value of $(a,b)$, the chiral superfields on
$M^{7|8}_{(a,b)}$ may be found which reproduce the semidirect type global
supersymmetry laws $\tilde\d$ for its $\bar M^{7|4}$ components. Then there should
exist a manifestly supersymmetric action.

\vskip 3mm\noindent
{\bf Acknowledgements} \\
The authors thank S.M. Kuzenko and M.A. Vasiliev for useful discussions.
The research of A. Yu. Segal is supported by grants INTAS-RFBR 95-829,
RFBR 98-02-16261. The research of A. G. Sibiryakov is partially supported
by grant RFBR 99-02-16617.

\setcounter{equation}{0}
\renewcommand{\theequation}{A.\arabic{equation}}
\vspace{5mm}
\noindent
{\bf\large APPENDIX A. Covariant derivatives on a coset space.}\\
Let us consider a group $G$, its subgroup $H$ and the homogeneous space
$M=G/H$. The group $G$ can be treated as a principle fibre bundle
(G,M,H) with the base space $M$ and the structure group $=$ standard fibre $H$.
This is the fibre bundle of the group in its cosets. All the covariant
derivatives used in the present paper arise in a construction
of this type.
Here we give general formulas and argue the existence of covariant
derivatives.

Let $X_a$, $a=1,\ldots,\dim H$ be a basis in Lie algebra $L_H$ of subgroup
$H$. The first supposition we have to do is that the set of vectors $X_a$
can be complemented by vectors $X_\a$, $\a=1,\ldots,\dim G - \dim H$ to
give the basis $\{X_a, X_\a\}$ in Lie algebra $L_G$ of $G$ so that the
commutation relations are partially graded:

\bea
[X_a,X_b] &=& C^c_{ab} X_c, \qquad
(L_H \mbox{~is subalgebra,}) \label{p-1}\\
{}[X_a,X_\a] &=& C^\b_{a\a} X_\b, \qquad C^b_{a\a} = 0,  \label{p0}\\
{}[X_\a,X_\b] &=& C^\g_{\a\b} X_\g + C^a_{\a\b} X_a.
\label{p1}\eea
Note that this choice of vectors $X_\a$ can always be done if $L_G$ is a
semisimple algebra and $L_H$ do not contain null vectors. Then $X_\a$ can
be chosen in to belong to orthogonal complement of $L_H$. For the factor-spaces
considered in this work the choice with (\ref{p0})
is always possible. But if we
chose the (Abelian) subalgebra spanned by $D^+_{\pm a}$ in the
superalgebra $\os$ (\ref{110}--\ref{1-10},\ref{13c}--\ref{18c}),
a choice with (\ref{p0}) would be impossible. This
could cause problems in attempts to construct covariant derivatives.

    Now we turn to the principle bundle (G,M,H). Let $\p: G\ra M$
be canonical projection
\be
\forall g \in G,\quad \forall h \in H \qquad
\p(gh)=gH \equiv m \in M
\label{p2}\ee
where left coset $gH$ is to be treated as a point of manifold $M$. A fibre
bundle is defined by trivializations  for
every open domain $U$ from an open covering of base space $M$.
Trivialization is a one-to-one map $\j$ satisfying
$$
\j:U\times H \ra \p^{-1}(U)\subset G, \qquad
\p \circ \j (m,h) = m.
$$
In our case this trivialization reduces to a smooth function $g(m)$ which
maps $U\ra G$ and satisfies $\p(g(m)) = m$. Then $\j(m,h) = g(m)h$.

   Let ${h^i}_j$ be matrices of a representation of subgroup $H$ in a
vector space $V$ in some basis:
$$
[hv]^i = {h^i}_j v^j,
$$
where $v^i$ are components of a vector $v\in V$. It is known that there is
the unique vector bundle $(E,H,V,H)$ associated with principle bundle
$(G,M,H)$ and the chosen representation of subgroup $H$ in a standard fibre
$V$. Let $\p_E$ be canonical projection in $E$ which maps the fibre $V_m$ over
the point $m$ to $m$: $\p_E(V_m) = m$. Sections $s(m)$ in the vector
bundle can be described by vector functions $v:G\ra V$ constrained by
\be
{h^i}_j v^j(gh) = v^i(g) \qquad \forall g\in G, \quad h\in H.
\label{p3}\ee
The bases in fibres $V_m = \p_E^{-1} (m)$ are defined by the
trivialization $g(m)$. The components of the section corresponding to the
function $v(g)$ (\ref{p3}) in these bases look like
\be
s^i(m) = v^i(g(m)).
\label{p4}\ee
The left action of the group $G$ on its homogeneous manifold is defined as
$$
\forall f \in G \qquad \cD_f:M\ra M, \qquad \cD_f(gH) = fgH, \qquad
\cD_f m = \p(fg(m)).
$$
This action induces the regular representation of the group $G$ in the
space of scalar functions $\f$ in the manifold $M$:
$$
T_f\f(m) = \f(\cD_{f^{-1}} m).
$$
Analogously the regular representation of the group $G$ in the space of
sections in the vector bundle $E$ can be defined. It is easier to do in
terms of vector functions $v$ (\ref{p3},\ref{p4}):
\be
T_f v(g) = v(f^{-1} g).
\label{p5}\ee
It is evident that new function $T_f v$ also satisfies the equation
(\ref{p3}) and
thus defines some section. We can write this action in the infinitesimal
form: for every vector $X \in L_G$
\be
T_X v(g) = l v(g)
\label{p6}\ee
here $l$ is right-invariant vector field corresponding to X,
\be
l(g) = Xg \equiv dR_g X, \qquad R_g g' =g' g
\label{p7}\ee
where $dR_g$ is the differential of the right shift $R_g$ (a right-invariant
vector field generates left action and therefore denoted '$l$').

    Let us denote $l_a$, $l_\a$ right-invariant and $r_a$, $r_\a$
left-invariant vector fields corresponding to basis vector fields $X_a$,
$X_\a$:
\be
\ba{ccc}
l_\a(g) = X_\a g, & l_a(g) = X_a g, & Xg \equiv dR_g X \\
r_\a(g) = g X_\a, & r_a(g) = g X_a, & gX \equiv dL_g X
\ea
\label{p8}\ee
where $L_g g' = gg'$. It is convenient to introduce also dual basis of
1-forms
$$
l^\a \cdot l_\b = \d^\a_\b, \qquad l^a \cdot l_b = \d^a_b, \qquad
l^\a \cdot l_b = l^a \cdot l_\b = 0
$$
and the same for $r^\a$, $r^a$ via $r_\a$, $r_a$ . Dots stand for the
contraction of vector with 1-form. Then, in every trivialization, the
connection in the principle bundle is the Lie algebra valued 1-form in the base space
being equal to
\be
A=(g^* l^a) X_a
\label{p9}\ee
where $g^*l^a$ is the pull-back of the 1-form $l^a$ from point $g(m)$ to point
$m$. The exterior covariant derivative is equal to
\be
\cD = d+(g^*l^a)X_a.
\label{p10}\ee
To obtain the usual covariant derivative the vielbein vector field $e_\a$
and its inverse 1-form have to be introduced in the base space:
\be
e^\a = g^* r^\a, \qquad e_\a = d\p (r_\a(g(m))),
\label{p11}\ee
here $d\p$ is the differential of the canonical projection. One can show that
$e^\a \cdot e_\b = \d^\a_\b$. Then the covariant derivatives are
\be
\cD_\a = e_\a \cdot \cD = e_\a + {A_\a}^a X_\a, \qquad
{A_\a}^a = e_\a\cdot(g^* r^\a).
\label{p12}\ee

   Consider a section $s$ in a vector bundle with a representation $T$ of
group $H$ on a fibre. Then $\cD_\a s$ is also a section in the vector
bundle associated with the same principle bundle and with the
representation $T\otimes \mbox{Ad}_\bot$ in a fibre. Here $\mbox{Ad}_\bot$
denotes the restriction of adjoint representation of subgroup $H$ to the
linear envelope of vectors $X_\a$. Further, one can show that
\be
\cD_\a s^i(m) = r_\a v^i (g(m)),
\label{p13}\ee
the action of covariant derivative on the section $s$ equals to the action
of left-invariant vector field on the function $v^i(g)$ (\ref{p4}).
That is why
derivatives $\cD_\a$ commute with left infinitesimal action (\ref{p6}) of the
algebra $L_G$ on sections. In this sense these derivatives are covariant.
The action of local rotations $M_a$ on a section can be generated by the
action of vector fields $r_a$ on the corresponding function $v(g)$ in the
group as follows from infinitesimal form of Eq.(\ref{p3}). Hence the
algebra of covariant derivatives and local rotations coincides with the
algebra of left-invariant vector fields, that is with algebra $L_G$:
\bea
[M_a,M_b] &=& C^c_{ab} M_c,  \\
{}[M_a,\cD_\a] &=& C^\b_{a\a} \cD_\b,   \\
{}[\cD_\a,\cD_\b] &=& C^\g_{\a\b} \cD_\g + C^a_{\a\b} M_a.
\label{p14}\eea

    Let us emphasize that the covariant derivatives $\cD_\a$ were
expressed for general case in terms of geometrical objects (left and right
invariant vector fields, projections and trivialization map g(m)) that are
available for any group $G$ and its subgroup $H$ provided in (\ref{p0})
$C^b_{a\a}=0$.  If we have two subgroups $H\subset L\subset G$, then three
factor-spaces arise: $G/H$, $G/L$ and $L/H$. It can be shown in general
case that covariant derivatives in space $G/H$ can be expressed in terms
of derivatives in space $L/H$, vector fields in $L$, projection $\p:L\ra
L/H$ and trivialization $l:L/H\ra L$. For short, we do not consider this
in detail. In subsection 3.a, we find these expressions for $G/H$ covariant
derivatives in particular case $G=\Osp$, $L=SL(2,C)$, $H=SU(2)$ from the
requirement that they satisfy the proper algebra. All the necessary
geometrical objects constructed from vector fields, projection and
trivialization in this case reduce to the 'small vielbeins' (\ref{d5}),
their tensor products and derivatives.

\setcounter{equation}{0}
\renewcommand{\theequation}{B.\arabic{equation}}
\vspace{5mm}
\noindent
{\bf\large APPENDIX B. The supertensor notation.}\\
Given a basis in a finite-dimensional supervector space $H$, a
supervector is characterized by a set of components $\eta_{\m}$, with
the Grassmann parity being assigned by the rule $\varepsilon
(\eta_{\m}) =\varepsilon (\m) \equiv \m$, where $\varepsilon (\m)
= 0,1 \,mod \,2$  is the grading function of this basis. Equivalently,
one has
\be  \label{a31}
\eta_{\mu_1} \eta_{\mu_2} = (-)^{\mu_1 \mu_2} \eta_{\mu_2} \eta_{\mu_1}
\ee
Let
\be
\eta'_{\m} = {G_{\m}}^{\n} \eta'_{\n} \,; \,
\varepsilon ({G_{\m}}^{\n}) = \varepsilon (\m) + \varepsilon (\n)
\equiv \m +\n
\ee
be a representation of a supergroup $G$.
The contragredient representation is defined by the rule
\be
\c'^{\m} = \c^{\n} {(G^{-1})_{\n}}^{\m} ,
\ee
where
\be
{(G^{-1})_{\m}}^{\r} {G_{\r}}^{\n} = {\d_{\m}}^{\n}
\ee
and ${\d_{\m}}^{\n}$ is an ordinary $\d$-symbol.

An invariant contraction, therefore, takes the form
\be \label{a35}
\c^{\n} \eta_{\nu} = (-)^{\nu} \eta_{\nu} \c^{\n} = Str \;\eta_{\nu} \c^{\m}
\ee
A supertensor $A_{\m_1 \m_2}{}^{\n_1}{}_{\m_3}{}^{\n_2}{}_{\dots}{}^{\dots}$
is defined to be transformed, by definition, as a product of supervectors and
supercovectors being multiplied in the same order the indices $\m_1 \,
\, \m_2 \, \n_1 \, \m_3 \, \n_2$ are :
\be
A_{\m_1 \m_2}{}^{\n_1}{}_{\m_3}{}^{\n_2}{}_{\dots}{}^{\dots}  =
{\eta_1}_{\m_1} \, {\eta_2}_{\m_2}\,
\c_3^{\n_1} \, {\eta_3}_{\m_3} \,{\c_2}^{\n_2} \dots
\ee
For example,
\be  \label{ab4}
{D'^{\n}}_{\m}  =(-)^{\n' (\n' + \n +\m' +\m)}
{(G^{-1})_{\n'}}^{\n} {G_{\m}}^{\m'}
{D^{\n'}}_{\m'},
\ee
To perform a covariant contraction (i.e., a contraction with
a supertensor result) over one upper and one lower index one have
to transfer them to each other by indices permutation and multiply the whole
expression by the corresponding sign factor according to the Eq.(\ref{a31})
and then contract them by the rule (\ref{a35}).
For example, one can make the contractions like
\be
\ba{l}
B_{\m_1 \m_2}{}^{\n_2}{}_{\dots}{}^{\dots} =
A_{\m_1 \m_2}{}^{\n}{}_{\n}{}^{\n_2}{}_{\dots}{}^{\dots}\\ \\
D_{\m_1}{}^{\n_1}{}_{\m_3}{}_{\dots}{}^{\dots}  =
(-)^{\n(\n_1 + \m_3 +1)} A_{\m_1
\n}{}^{\n_1}{}_{\m_3}{}^{\n}{}_{\dots}{}^{\dots} \\ \\
B_{\m_1}{}_{\dots}{}^{\dots} =
(-)^{\m} B_{\m_1 \m}{}^{\m}{}_{\dots}{}^{\dots} =
(-)^{\m} A_{\m_1 \m}{}^{\n}{}_{\n}{}^{\m}{}_{\dots}{}^{\dots}
\ea
\ee
Two nonzero tensors may be equal only if their indices orders are equivalent.
Since we will work with tensor expressions in the main text, the
indices order in the right hand side of some equation is unambiguously
determined by the one of the left hand side. With this in mind we equivalently
represent the formulas like (\ref{ab4}) as follows
\be  \label{ab5}
{D'^{\n}}_{\m} \; [=]  \;
{(G^{-1})_{\n'}}^{\n} {G_{\m}}^{\m'}
{D^{\n'}}_{\m'}
\ee
The notation "$[=]$" means that each term in the right hand side should
be multiplied by a sign factor arising when the indices to
be contracted are transferred to their partners and the rest indices are
put to the same order as in the left hand side. It's easy to see that
this prescription determines the sign factors unambiguously.

The supersymmetrization  and superantisymmetrization  are defined standardly
as follows,
\be \label{ab6}
\ba{l}
A_{(\m\n)} = \frac{1}{2} \left( A_{\m\n} +(-)^{\mu \nu} A_{\n\m}\right) \\ \\
A_{[\m\n]} = \frac{1}{2} \left( A_{\m\n} -(-)^{\mu \nu} A_{\n\m}\right),
\ea
\ee
the generalization to the multiple number of indices is straightforward.
We also employ the condensed notation $A_{\mu
(n)}$ and $B_{\mu[n]}$ for supersymmetric ands superantisymmetric tensors of
rank $n$.

The  standard notation
\be
(-)^{\e(k)  \m(l) }  \equiv  (-)^{\e[k]  \m[l] }  \equiv (-)^{(\e_1 +\e_2
+\dots e_k)(\m_1 +\m_2  + \dots m_l)} \ee is assumed.

In the main text, one meets the supertensors transforming in the tensor products
of two nonequivalent representations. All the rules of this appendix are
assumed to be employed, the only point is that the signs factors
arising from a permutation of nonequivalent indices should be also taken
into account.

\newpage
\setcounter{equation}{0}
\renewcommand{\theequation}{C.\arabic{equation}}
\vspace{5mm}
\noindent
{\bf\large APPENDIX C. The $A(1|0)$ superalgebra and
properties of its finite
dimensional representations.}\\
The $A(1|0)$ superalgebra is defined via the following graded commutation
relations:
\be \label{a21}
\ba{c} \label{A21}
{} [ S_{a(2)}, S_{b(2)} ] = 2 \varepsilon_{ab} S_{ab} \\  \\
{}[ S_\as, D_{\pm b} ] = \varepsilon_{ab} D_{\pm a} \\  \\
{}[ T, D_{\pm a} ] = \pm \frac{1}{2} D_{\pm a} \\   \\
{}[ D_{+a} , D_{-b} ] =S_{ab}+\ve_{ab}T \; \; ; \; \; S_{ab}=S_{ba}
\ea
\ee
The other commutators vanish.
Among the real forms of this superalgebra is $su(2,0|1,0)$ superalgebra with
the even part being isomorphic to $su(2) \oplus so(2)$, this real form
is explored in the main text.

The irreducible
finite dimensional representations (fdrs) of $A(1|0)$ were classified by
Tierry-Mieg \cite{tmieg}. For our purposes it is useful, however, to develop
an independent treatment, including non completely reducible fdrs. We do not
pursue an exhaustive classification and just describe the fdrs relevant for the
constructions of the paper and their close relatives, along with the
information appropriate for our exposition in the main text.

First, let's find the representations with smallest dimensions.
Since the even generators of $A(1|0)$ are expressed via the odd ones
(see Eq.(\ref{A21})),
a fdr is trivial if the odd generators act trivially. Therefore,
there are no nontrivial $(1|0), (0|1) , (1|1)$ representations.
For the dimension $(1|2)$, one finds exactly two inequivalent fdrs:
$(1|2) \equiv \x$ and $\overline{(1|2)} \equiv \xb$, they are both
irreducible.  We
denote the elements of corresponding supervector spaces by $\f_{\x}$ and
$\f_{\xb}$.  The decomposition of
$\f_{\x}, \f_{\xb}$ representations w.r.t. even subalgebra $su(2) \oplus so(2)$
and the action of the odd generators in such a basis looks as follows
\be \label{a22} \label{A2x}
\ba{cc}
\f_{\x} = \left( \ba{c} \f_{+a} \\ \f_{++} \ea \right) ;&
\f_{\xb} = \left( \ba{c} \f_{-a} \\ \f_{--} \ea \right)
\ea
\ee
\be \label{a23}
\ba{cccc}
D_{+b} \left( \ba{c} \f_{+a} \\ \f_{++} \ea \right) =&
\left( \ba{c} \varepsilon_{ba} \f_{++} \\ 0 \ea \right) \hspace{5mm}; &
D_{+b} \left( \ba{c} \f_{-a} \\ \f_{--} \ea \right) =&
\left( \ba{c} 0 \\ \f_{-b} \ea \right) \\ & & & \\
D_{-b} \left( \ba{c} \f_{+a} \\ \f_{++} \ea \right) =&
\left( \ba{c} 0 \\ \f_{+b} \ea \right) \hspace{5mm}; &
D_{-b} \left( \ba{c} \f_{-a} \\ \f_{--} \ea \right) =&
\left( \ba{c} \varepsilon_{ba} \f_{--} \\ 0 \ea \right)
\ea
\ee
Here the pluses and minuses label the $T$-weights, namely,
\be \label{a24}
T \f_{\pm a} = \frac{1}{2} \f_{\pm a}\;\; ; \;\;T \f_{\pm \pm} = \pm \f_{\pm \pm}
\ee
It is easy to see by direct analysis of commutation relations
(\ref{A21}) that these $T$-weights combinations are the only ones allowing to
construct a nontrivial $(1|2)$ dimensional $A(1|0)$ representation.
In what follows, we adopt a simpler notation for the  indices running
the fdr's basis:  $\xi =(z,a) ; {\bar \xi} = ({\bar
z},{\bar a}) $, where $z, {\bar z}$ label one-dimensional even subspaces.

Two representations, $\x$ and $\xb$, are conjugated in the sense
that they are mapped to each other by an outer $A(1|0)$-automorphism of the
form
\be \label{a25}
T' = -T \;\;,\;\; {S'}_{ab} = S_{ab}\;\;,\;\; {D'}_{\pm a} = {D}_{\mp a}
\ee
As a consequence, there exist the second rank invariant tensors $C_{\e \zb}$
and $C_{\eb \z}$ (see App. B for our supertensor notation):
\be \label{a26}
\ba{cc}
C_{\e \zb} =\left( \ba{cc} -\varepsilon_{ab} & 0 \\ 0 & 1 \ea \right) \hspace{5mm}; &
C_{\eb \z} = (-)^{\eb \z}  C_{\z \eb} =
\left( \ba{cc} -\varepsilon_{ab} & 0 \\ 0 & 1 \ea \right),
\ea
\ee
as well as their inverse tensors $C^{\eb \z}$ and $C^{\e \zb}$  :
\be \label{a27}
\ba{cc}
C^{\e \zb} =\left( \ba{cc} \varepsilon^{ab} & 0 \\ 0 & 1 \ea \right) \hspace{5mm}; &
C^{\eb \z} = (-)^{\eb \z}  C_{\z \eb} =
\left( \ba{cc} \varepsilon^{ab} & 0 \\ 0 & 1 \ea \right),
\ea
\ee
satisfying the relations
\be \label{a28}
\ba{c}
C^{\eb \z} C_{\z \etab} = (-)^{\eb} {\d^{\eb}}_{\etab}\;\;;\;\;
(-)^{\zb} C_{\e \zb} C^{\zb \eta} =  {\d_{\e}}^{\eta} \\ \\
C^{\eb \z} C_{\z \eb} = C^{\e \zb} C_{\zb \e} = -1 \\ \\
\ve(C^{\eb \z}) = \ve(\eb) + \ve(\z) =0,
\ea
\ee
where ${\d^{\eb}}_{\etab}$ and ${\d_{\e}}^{\eta}$ are the ordinary
$\d$-symbols. These tensors may be used to raise and lower the $\x$ and $\xb$
indices by the rule:
\be \label{a29}
\ba{cc}
\f^{\e} = C^{\e \eb} \f_{\eb} \;\; ; &
\f^{\eb} = C^{\eb \e} \f_{\e} ;  \\
\f_{\e} = \f^{\eb} C_{\eb \e}  \;\; ; &
\f_{\eb} = \f^{\e} C_{\e \eb}
\ea
\ee
Moreover, these invariant supermatrices enable one to obtain tensor
form of expressions (\ref{a22}--\ref{a24}) for the action of the
generators of $A(1|0)$ on $\f_\e$ and $\f_\eb$:
\bea
\cE_{\eb\z} \f_\etab &=& C_{\eb\z} \f_\etab + \f_\eb C_{\z\etab}, \cr
\cE_{\eb\z} \f_\eta &[=]& -(C_{\eb\z} \f_\eta + C_{\eb\eta} \f_\z).
\non
\eea
To discuss tensor products of $\x,\xb$, it's worth working out $(2|2)$
dimensional representations. Obviously, their decomposition w.r.t. even
subalgebra should contain two $su(2)$ scalars in the even sector and
one spinor in the odd sector, assigned with corresponding $T$-weights
(otherwise the $(2|2)$ is a direct sum of fdrs with lower dimensions).

Direct analysis of commutation relations
(\ref{a21})
shows that there is the unique series
of $(2|2)$ fdrs, which we denote $(2|2)_p$, parametrized by a complex number
$p \in \bf C$. Denote the corresponding index by $i_p$. The decomposition w.r.t
even subalgebra looks as
\be \label{a210} \label{A25}
\ba{cc}
\f_{i_p} = \left( \ba{c} \f_{pa} \\ \f_{p+1} \\ \f_{p-1} \ea \right); &
T \left( \ba{c} \f_{pa} \\ \f_{p+1} \\ \f_{p-1} \ea \right) =
 \left( \ba{c} \frac{p}{2} \f_{pa} \\ \frac{p+1}{2} \f_{p+1} \\
 \frac{p-1}{2} \f_{p-1} \ea \right) ,
\ea
\ee
while the odd generators action is set by the formulas
\be \label{a211}
\ba{cc}
D_{+b} \left( \ba{c} \f_{pa} \\ \f_{p+1} \\ \f_{p-1} \ea \right) =
\left( \ba{c} \varepsilon_{ba} \f_{p+1} \\ 0 \\ \frac{1-p}{2} \f_{pb} \ea \right) ; &
D_{-b} \left( \ba{c} \f_{pa} \\ \f_{p+1} \\ \f_{p-1} \ea \right) =
\left( \ba{c} \varepsilon_{ba} \f_{p-1} \\ \frac{1+p}{2} \f_{pb} \\ 0 \ea
\right)
\ea
\ee
The automorphism (\ref{a25}) maps the $(2|2)_p$ and $(2|2)_{-p}$ fdrs onto
each other.  The corresponding second rank invariant tensor $C_{i_p j_{-p}}$
has the form
\be \label{a212} C_{i_p j_{-p}}   = \left( \ba{ccc}
-\varepsilon_{ab} & 0 & 0 \\ 0  & 0 & \frac{1+p}{2} \\ 0  & \frac{1-p}{2} & 0
\ea \right) .
\ee
The $(2|2)_p$ fdrs are irreducible for any $p$ except the values $p=\pm 1$.
In those cases, the fdrs $(2|2)_1$ and $(2|2)_{-1}$ have the structure of
semidirect product $(2|2)_{1}  = (1|0) \rightarrow \x$, $(2|2)_{-1}  =
(1|0) \rightarrow \xb$  with invariant subspaces carrying the
$(1|2)$ dimensional fdrs and the trivial representation being realized in the
factor space.

Obviously, the $T$-weights structure appearing in the decomposition
(\ref{A25}) is the only possible one for $(2|2)$ dimensional fdrs.
The representations $(2|2)_p$
exhaust all $(2|2)$ fdrs in the case $p \neq \pm 1$. In the last two cases. one
finds two additional representations contragredient to  $i_{1}$ and
$i_{-1}$ : $\tilde{i}_1$ and $\tilde{i}_{-1}$.
Their structure is as follows:
\be \label{a213}
\ba{cc}
\j_{\tilde{i}_1} = \left( \ba{c} \j_{-a} \\ \j_{-2} \\ \j_{0} \ea \right);
& \j_{\tilde{i}_{-1}} = \left( \ba{c} \j_{+a} \\ \j_{2} \\ \j_{0} \ea
\right)
\ea
\ee
the odd generators action is
\be \label{a214}
\ba{cc}
D_{+b} \left( \ba{c} \j_{+a} \\ \j_{2} \\ \j_{0} \ea \right) =
\left( \ba{c} \varepsilon_{ba} \j_{2} \\ 0 \\ \j_{+b} \ea \right) ; &
D_{-b} \left( \ba{c} \j_{+a} \\ \j_{2} \\ \j_{0} \ea \right) =
\left( \ba{c} 0 \\ \j_{+b} \\ 0 \ea \right)
\ea
\ee

\be \label{a215}
\ba{cc}
D_{-b} \left( \ba{c} \j_{-a} \\ \j_{-2} \\ \j_{0} \ea \right) =
\left( \ba{c} \varepsilon_{ba} \j_{-2} \\ 0 \\ \j_{-b} \ea \right) ; &
D_{+b} \left( \ba{c} \j_{-a} \\ \j_{-2} \\ \j_{0} \ea \right) =
\left( \ba{c} 0 \\ \j_{-b} \\ 0 \ea \right)
\ea
\ee
These two representations, along with the direct sums
$\x \oplus (1|0)\; ; \; \xb \oplus (1|0) \;;\;
(1|0) \oplus (1|0) \oplus (0|1) \oplus (0|1)$ complete the list of all
$(2|2)$ dimensional fdrs.

Further, let us show that the supersymmetric tensor
product of $n,\,(n \neq 1)$ $\x$ representations is isomorphic to
$(2|2)_{2n-1}$ and analogously for $\xb$ (the round brackets denote the
supersymmetrization):
\be \label{a216} \label{A26}
(\x_1 \dots \x_n ) = (2|2)_{2n-1} \;\;,\;\;
(\xb_1 \dots  \xb_n ) = (2|2)_{-(2n-1)}.
 \ee
Indeed, consider tensor $\f_{(\x_1 \dots \x_n)}$.
It's independent nonzero components are
\be \label{a217}
\ba{c} \f_{z,z,\dots,z}\equiv \f_{2n} \\
\f_{a,z,\dots,z} \equiv \f_{2n-1\, a} \\
\f_{a,b,z,\dots,z} \equiv \varepsilon_{ab} \f_{2n-2}
\ea \ee
As the component content of $\f_{(\x_1 \dots \x_n)}$ is identical to that
of $(2|2)_{2n-1}$ and there is no another $(2|2)$ dimensional representation
with such $T$-weight structure, these representations are isomorphic. Of
course, this assertion may be proved by explicit calculation either.

The next important fact we would like to deliver deals with the
supersymmetrized product of two $(2|2)_p$ fdrs.
\be \label{a218}  \label{A27}
((2|2)_p \otimes (2|2)_p ) = (2|2)_{2p-1}
\oplus (2|2)_{2p+1} \;\;, p\neq 0
\ee
It is proved by direct calculation, so we do not dwell on detailes here.
Making use the Eq.(\ref{A26}), one can rewrite (\ref{A27}) in the case $p
\in {\bf Z_+}$ in the manner
(the external round brackets in the l.h.s. stand for the supersymmetrization
w.r.t.  two groups of indices, $\xi$ and $\eta$, and not the complete
supersymmetrization over all indices)
\be \label{a219}
((\x_1 \dots \x_n) \otimes (\eta_1 \dots \eta_n)) =
(\e_1 \dots \e_{2n-1}) \oplus (\e_1 \dots \e_{2n})\;\; ; n\neq 1.
\ee
Equivalently, given two supersymmetric rank-$n$ tensors, their sypersymmetric
tensor product is decomposed as follows
\be \label{a220} \ba{c} \label{A28}
\f_{\x_1 \dots \x_n}
\j_{\eta_1 \dots \eta_n} -(-)^{\x(n)\eta(n)}
\f_{\eta_1 \dots \eta_n} \j_{\x_1 \dots \x_n} =\\  \\ =
\f_{(\eta_1 \dots \eta_n }
\j_{\x_1 \dots \x_n)} +
{C_{\x_1 \dots \x_n;\eta_1 \dots \eta_n}}^{\e_1 \dots \e_{2n-1}}
\c_{\e_{2n-1} \dots \e_1} ; n\neq 1.
\ea
\ee
We see that there exist two subspaces in the supersymmetrized product of two
sypersymmetric tensors: one is the complete supersymmetrization over all
indices, and  $(\e_1 \dots \e_{2n-1}) $ is something new. Moreover, we see that
there exists an invariant $4n-1$-rank tensor
${C_{\x (n);\eta (n)}}^{\e(2n-1)}$.

It follows from (\ref{A27}) that
${C_{\x (n);\eta (n)}}^{\e(2n-1)}$ has the properties
\be \label{a221} \label{A211'}
\ba{c}
{C_{\x (n);\eta (n)}}^{\e(2n-1)}   = (-)^{\x (n)\eta (n)} \,
{C_{\eta (n);\x (n)}}^{\e(2n-1)} \\ \\
{C_{\x_1 ... (\eta;\eta_1 .... \eta_n)}}^{\e(2n-1)}=0 \\ \\
C^{{\bar \e}\z} C_{\x_1 \dots \x_n;\eta_1 \dots \eta_{n-1} \z;{\bar\e}
{\bar \e_2}\ldots {\bar\e_{2n-1}}} =0  \\  \\
\ve({C_{\x (n);\eta (n)}}^{\e(2n-1)}) =\x (n) +\eta (n) +\e(2n-1) =0.
\ea
\ee

Obviously, the higher rank $C$-tensors may be expressed via the
lower rank ones in the manner
\be \label{a222}
{C_{\x (n+1);\eta (n+1)}}^{\e(2(n+1)-1)}
[=] {C_{\x(n);\eta(n)}}^{\e (2n-1)} {\d_{\xi}}^{\e} {\d_{\eta}}^{\e}.
\ee

The lowest rank tensor in this sequence is
$ {C_{(\x_1 \x_2);(\eta_1 \eta_2) ;(\eb_1 \eb_2 \eb_3)}}$ .
Explicitly, it has the components
\be \label{a223} \label{A29}
\ba{c}
C_{ab\,;\,cd\,;\, \bar{z}\, \bar{e}\, \bar{f}} =
\varepsilon_{ab} \varepsilon_{cd} \varepsilon_{ef} \qquad
C_{ab\,;\,cz\,;\,\bar{z} \, \bar{z} \, \bar{f}} = \varepsilon_{ab}
\varepsilon_{cf} \\ \\
C_{az\,;\,bz\,;\, \bar{z} \, \bar{z}\, \bar{z}} = \varepsilon_{ab} \qquad
C_{ab\,;\,zz\,;\, \bar{z} \, \bar{z} \, \bar{z}} = -2\varepsilon_{ab} \\  \\
\ea
\ee
and those obtained from (\ref{A29}) by indices permutations, the other
components equal zero. The invariance of tensor (\ref{a223}) is checked by
direct calculation.

Introduce the next important tensor
$ {{\tilde C}_{[\x_1 \x_2];[\eta_1 \eta_2] ;(\eb_1 \eb_2 \eb_3)}}$ which is
superantisymmetric in the first two index pairs and supersymmetric
in the last three-index group:
\be \label{a224} \label{A210}
\ba{l}
{\tilde C}_{[\x_1 \x_2];[\eta_1 \eta_2] ;({\bar \e_1}{\bar \e_2}{\bar
\e_3})}
= (-)^{(\x_1 +\x_2)\eta_1}
{C}_{\eta_1 \x_1;\x_2 \eta_2 ;({\bar \e_1}{\bar \e_2}{\bar
\e_3})} - \\  \\
-(-)^{\x_1 \x_2 +(\x_1 +\x_2)\eta_1}
{C}_{\eta_1 \x_2;\x_1 \eta_2 ;({\bar \e_1}{\bar \e_2}{\bar
\e_3})} .
\ea
\ee
The inverse transformation reads
\be \label{a225}
\ba{l}
{C}_{(\x_1 \x_2);(\eta_1 \eta_2) ;({\bar \e_1}{\bar \e_2}{\bar
\e_3})}
=  -\frac{1}{3} \left( (-)^{(\x_1 +\x_2)\eta_1}
{\tilde C}_{[\eta_1 \x_1];[\x_2 \eta_2] ;({\bar \e_1}{\bar \e_2}{\bar
\e_3})} \right. + \\ \\
+ \left. (-)^{\x_1 \x_2 +(\x_1 +\x_2)\eta_1}
{\tilde C}_{[\eta_1 \x_2];[\x_1 \eta_2] ;
({\bar \e_1}{\bar \e_2}{\bar \e_3})} \right)
\ea
\ee
The $\tilde C$ tensor is also supersymmetric w.r.t. permutation of
two first groups of indices, and traceless:
\be \label{a226}  \label{A211}
\ba{c}
{\tilde C}_{[\x_1 \x_2];[\eta_1 \eta_2] ;
({\bar \e_1}{\bar \e_2}{\bar \e_3})}
=   (-)^{(\x_1+\x_2)(\eta_1 +\eta_2) }
{\tilde C}_{[\eta_1 \eta_2];[\x_1 \x_2];
({\bar \e_1}{\bar \e_2}{\bar \e_3})}
\\  \\
C^{{\bar \e}\z} {\tilde C}_{[\x_1 \x_2] ;[\eta_1 \z];
({\bar \e}{\bar \e_2}{\bar \e_3})} =0  ,
\ea
\ee
and satisfies the property
\be \label{a227} \label{A212}
\ba{c}
{\tilde C}_{\x_1 [\eta_1; \eta_2 \eta_3] ;
({\bar \e_1}{\bar \e_2}{\bar \e_3})} =0  ,
\ea
\ee
The component content of the $\tilde C$ tensor is
\be \label{a228}
\ba{c}
{\tilde C}_{ab\,;\,cd\,;\,\bar{z}\, \bar{e}\, \bar{f}} = (\varepsilon_{ca}
\varepsilon_{bd} + \varepsilon_{cb} \varepsilon_{ad})\varepsilon_{ef} \qquad
{\tilde C}_{ab\,;\,cz\,;\,\bar{z} \,\bar{z} \,\bar{f}} = \varepsilon_{ca}
\varepsilon_{bf}+ \varepsilon_{cb} \varepsilon_{af} \\  \\ {\tilde
C}_{az\,;\,bz\,;\,\bar{z} \,\bar{z}\, \bar{z}} = \varepsilon_{ab} \qquad
{\tilde C}_{ab;zz;\bar{z} \bar{z} \bar{z}} = 0
\ea
\ee
The formulas
(\ref{A211'}, \ref{A211} \ref{A212}) just indicate that, for $C$ and $\tilde
C$, two first groups of indices are characterized by the Young tableaux of the
form

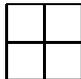
\unitlength=0.63mm \special{em:linewidth 0.4pt} \linethickness{0.4pt}
\begin{picture}(30.00,40.00)(-50.0,117.0)
\put(10.00,150.00){\line(1,0){16.00}}
\put(26.00,150.00){\line(0,-1){16.00}}
\put(26.00,134.00){\line(-1,0){16.00}}
\put(10.00,134.00){\line(0,1){16.00}}
\put(10.00,142.00){\line(1,0){16.00}}
\put(18.00,150.00){\line(0,-1){16.00}}
\end{picture}

The completely analogous formulas take place for conjugated representations,
therefore, we also have the conjugated tensors
${\bar C}_{(\xb_1 \xb_2);(\etab_1 \etab_2) ;(\e_1 \e_2 \e_3)}$,
${\bar {\tilde C}}_{[\xb_1 \xb_2];[\etab_1 \etab_2] ;(\e_1 \e_2 \e_3)}$,
just with indices $\x$ being substituted for $\bar \x$.
In what follows, we will not write the bar over the $C$-letters as it is
clear from the index structure which tensor is used.

The tensor $\tilde C$ is relevant for the component description of the
equality
\be \label{a229}
\left( \x[2] \otimes  \x[2] \right) = \e[4] \oplus
\e(3),
\ee
which is proved by direct analysis. We see that there are two
irreducible subspaces in the supersymmetric tensor product of two
second rank superantisymmetric tensors:  the first is obtained by complete
superantisymmetrization over all indices, the second one is extracted with
the $\tilde C$ help by the rule
\be \label{a230}
\ba{l}
r_{\e(3)} = \left( f^{[\etab_2 \etab_1]} g^{[\xb_2 \xb_1]} -(-)^{\xb[2]
\etab[2]} g^{[\xb_2 \xb_1]} f^{[\etab_2 \etab_1]} \right) {\tilde
C}_{\xb_1 \xb_2 ; \etab_1 \etab_2 ; \e(3)} .
\ea
\ee
The last formula plays an
important role in the main text
(see Eqs.(\ref{664})).

Let us eventually describe the structure of fdrs $I_p =\x(p) \otimes \xb(p)
$.
The cases $p=1$ and $p\neq 1$ are different.
For $p=1$, one finds
\be \label{a2}
\x \xb = \stackrel{\circ}{\x \xb}   \oplus (1|0)
\ee
where $\stackrel{\circ}{ \x \xb}$ stands for the traceless part of
$\x \xb$ fdr -- the adjoint fdr of $A(1|0)$ (see Eq.(\ref{117})),
and $(1|0)$ is the
trivial fdr. This assertion is easy to prove as every element of the $\x \xb$ fdr
possesses the unique decomposition
\be \label{a2222}
\f_{\x \xb} [=] \stackrel{\circ}{\f}_{\x \xb} - C_{\x \zb} {\f^{\zb}}_{\xb}
\;\;;\;\; {\stackrel{\circ}{\f}{}^{\zb}}_{\zb} =0.
\ee
It appears that all
$I_p, p \geq 2$ are isomorphic:  $\forall p \, \geq 2 \;\;I_p = F $ and have the
structure
\be \label{a231} \label{A2ip}
F = (1|0) \rightarrow \left(\x \oplus \xb \oplus
\stackrel{\circ}{\x \xb} \right) \rightarrow (1|0)
\ee
Let's prove this statement.
Exhibit the component content of the $\f_{\e(2)\eb(2)}$ supertensor:
\be \label{a232}
\f_{\e(2)\eb(2)}= \left\{ \f, \p, \c, \f_{+a}, \f_{-a}, \p_{+a}, \p_{-a},
\f_2, \f_{-2}, \p_{a(2)} \right\}
\ee
The isomorphism
$I_p = F \;\;\forall p \neq\pm1$ is established easily: one observes that
dimensions of all $I_p$ fdrs are $(8|8)$ and equal to each other, on the other
hand, there exist an
invariant map $I_{p+1}$ {\it on} $I_p$ which may be written in
tensor notation as follows:
\be \label{a233}
\f_{\e(p)\eb(p)} = C^{\zb \z} \f_{\e(p) \z \zb \eb(p)}.
\ee
So we may take $p=2$ without the loss of generality and represent a general
element of $F$ representation space by $\f_{\e(2)\eb(2)}$.
There exist the invariant subspace of double-traceless tensors:
\be \label{a234}
H^{7|8} : \{ \f_{\e(2)\eb(2)} \in  H^{7|8} | {\f^{\eb \zb}}_{\zb \eb} =0 \}
\ee
The factor space $F/H^{7|8}$ carries $(1|0)$ dimensional (trivial) fdr, so one
observes that $F = (1|0) \rightarrow H^{7|8}$.

Further, there exist the invariant subspace
(the "$[=]$" symbol stands for the ordinary
equality with an account of a sign
factor, see App.B)
\be \label{a235}
H^{1|0} \subset H^{7|8} : \{ \f_{\e(2)\eb(2)} \in  H^{1|0} | \f_{\e \e \eb \eb}
\;\;[=]\;\; C_{\e \eb} C_{\e \eb} \o \}
\ee
the inclusion $H^{1|0} \subset H^{7|8}$ is clear as the double-$C^{\e\eb}$ trace in
$H^{1|0}$ is proportional to the supertrace of the identity operator
in the $F$ space, which is equal zero as $F$ is $(8|8)$-dimensioinal.
Therefore, one gets the structure $H^{7|8} =(6|8) \rightarrow H^{1|0}$, where
$(6|8)$ is some $(6|8)$-dimensional fdr.
Consider two subspaces:
\be \label{a236}
\ba{c}
H_{trace} \subset H^{7|8} : \{ \f_{\e(2)\eb(2)} \in  H_{trace} |
\f_{\e \e \eb \eb} \;\;[=]\;\; C_{\e \eb} \f_{\e \eb} \,;\, \f_{\e \e \eb \eb}
\in H^{7|8} \} \\ \\
H_{traceless} \subset H^{7|8} : \{ \f_{\e(2)\eb(2)} \in  H_{traceless} |
{{\f_{\e}}^{ \zb}}_{\zb \eb} = 0 \} \\
\ea
\ee
These two subspaces, $H_{trace}$ and $H_{traceless}$, overlap only via the
$H^{1|0}$ since
\be \label{a237}
\ba{c}
\f_{\e(2)\eb(2)} \in  H_{trace} \Rightarrow
\f_{\e \e \eb \eb} = C_{\e \eb} \f_{\e \eb} , \\   \\
\f_{\e(2)\eb(2)} \in  H_{traceless} \Rightarrow
\f_{\e \eb} = C_{\e \eb} {\f^{\zb}}_{\zb} \Rightarrow
\f_{\e(2)\eb(2)} \in  H^{1|0}
\ea
\ee
It is easy to see that the component form of
$H_{trace}$ and $H_{traceless}$ is
\be \label{a238}
\ba{c}
H_{trace} = \left\{\c, \p, \p_{+a}, \p_{-a}, \p_{a(2)} \right\}\\ \\
H_{traceless} =  \left\{ \c, \f, \f_{+a}, \f_{-a}, \f_2, \f_{-2}, \right\}
\ea
\ee
and their dimensions are $(5|4)$ and $(3|4)$, correspondingly.
As they have $(1|0)$-dimensional overlap (parametrized by $\c$ in the
last formula),
one has $H^{7|8} =H_{trace} + H_{traceless}\;$ and
$\; H^{7|8}/H^{1|0} = (H_{trace} /H^{1|0}) \oplus (H_{traceless}/H^{1|0}$)
in the factor space. The representation
$H_{trace} /H^{1|0}$ is isomorphic to
$\stackrel{\circ}{\x \xb}$ as $\x \xb = \stackrel{\circ}{\x \xb} \oplus (1|0)$
(see Eqs.(\ref{a2}, \ref{a2222}))
while $H_{traceless} /H^{1|0}$ is some $(2|4)$-dimensional fdr
with component content exhibited by $\f$'s in the formula (\ref{a238}).
Obviously, it is decomposed into direct sum of two $(1|2)$-fdrs with
components $\{ \f_{+a}, \f_2 \}$ and $\{ \f_{-a}, \f_{-2} \}$, which are
identified with two unique representations with such dimension (see
Eq.(\ref{A2x})) :  $H_{traceless} /H^{1|0} =\x \oplus \xb$.

Therefore, we come to the overall result of our analysis in the form of
the Eq.(\ref{A2ip}). The fact that the decomposition (\ref{A2ip}) is not a
direct sum is checked by explicit component calculation.

Let's make a few remarks.

One may wonder about the explicit tensor form of the $H_{traceless}$
subspace. It is easy to prove that every element of $H_{traceless}$ may
be written in the form
\be \label{a239}
 \f_{\e(2)\eb(2)} \;\;[=]\;\;
 C_{\e(2);\z(2);\eb(2) \zb} \f^{\zb \z(2) }
+ C_{\eb(2);\zb(2);\e(2) \z} \f^{\z \zb(2) }
\ee
the invariant seven-rank $C$-tensors play an essential role here.
The last formula presents a general solution to the tracelessness constraint,
with the first and the second terms in the r.h.s. parametrizing the preimages
(w.r.t.  factorization by $H^{1|0}$) of $\xb$ and $\x$ fdrs, $H_{\xb}$
and $H_{\x}$, correspondingly.

One may impose another constraint in $H^{7|8}$:
\be \label{a240'}
0\;\;[=]\;\;{{C_{\e(2);}}^{\zb(2)} {}_{;\eb}}{}^{\z(2)} \f_{\z(2) \zb(2)}
\ee
It's now easy to derive that its general solution is
\be \label{a240}
\f_{\z(2) \zb(2)} \in H_{\x} + H_{trace}.
\ee
Quite analogously,
\be \label{a241}
\ba{c}
0\;\;[=]\;\;{{C_{\eb(2);}}^{\z(2)} {}_{;\e}}{}^{\zb(2)} \f_{\z(2) \zb(2)}
\Rightarrow \\ \\ \Rightarrow
\f_{\z(2) \zb(2)} \in H_{\xb} + H_{trace}.
\ea
\ee
The joint effect of constraints (\ref{a240'}) and (\ref{a241}) is $H_{trace}$
subspace.

\end{document}